\documentclass[%
 aip,
 amsmath,amssymb,
 reprint,%
]{revtex4-1}

\usepackage{comment}
\usepackage{csquotes}
\usepackage{amsthm}
\usepackage{svg}
\usepackage{graphicx}
\usepackage{dcolumn}
\usepackage{bm}
\usepackage{xcolor}
\usepackage{hyperref}
\usepackage{array} 
\usepackage{svg}
\usepackage{booktabs}    
\usepackage{amsmath}     
\usepackage{siunitx} 
\usepackage{xcolor}
\usepackage{svg}  
\usepackage{xcolor}
\usepackage{framed}
\usepackage{quantikz}
\newtheorem{theorem}{Theorem}
\newtheorem{definition}{Definition}
\newtheorem{identity}{{Definition}}

\newtheorem{statement}{Statement}
\newtheorem{corollary}{Corollary}[theorem]

\usepackage{multirow} 
\usepackage[caption=false]{subfig} 

\usepackage[T1]{fontenc}
\usepackage{amssymb}
\usepackage{algorithm}
\makeatletter
\@ifclassloaded{revtex4-2}{\edef\ftype@algorithm{\the\numexpr\c@float@type/2\relax}}{}
\@ifclassloaded{revtex4-1}{\edef\ftype@algorithm{\the\numexpr\c@float@type/2\relax}}{}
\makeatother
\usepackage{algpseudocode}

\newcommand{\stabilizerHamiltonian}{H_{\mathrm{S}}}
\newcommand{\commutingHamiltonian}{H_{\mathrm{C}}}
\newcommand{\anticommutingHamiltonian}{H_{\mathrm{A}}}
\newcommand{\noncontextualHamiltonian}{H_{\mathrm{NC}}}
\newcommand{\lineGraphFreeFermionHamiltonian}{H_{\mathrm{FF}}}
\newcommand{\freeFermionHamiltonian}{H_{\mathrm{FFD}}}

\newcommand{\Id}{\mathbb{I}}       

\newcommand{\ac}[2]{[#1, #2]_+}    
\newcommand{\comm}[2]{[#1, #2]}    

\newcommand{\Fn}[2]{\textsc{#1}(#2)}     
\newcommand{\True}{\mathtt{True}}        
\newcommand{\False}{\mathtt{False}}
\newcommand{\coeff}{f}
\newcommand{\supp}{\Gamma}               
\newcommand{\pool}{S}                    
\newcommand{\subdiag}{\mathrm{d}}        
\newcommand{\subquad}{\mathrm{q}}        
\newcommand{\subrest}{\mathrm{r}}        
\newcommand{\pooldiag}{\pool_{\subdiag}} 
\newcommand{\poolquad}{\pool_{\subquad}} 
\newcommand{\poolrest}{\pool_{\subrest}} 
\newcommand{\wpool}{W}                   
\newcommand{\seed}{Q}                    
\newcommand{\frag}{F}                    
\newcommand{\cand}{B}                    
\newcommand{\defer}{D}                   
\newcommand{\kLinIndepComm}{k}           
\newcommand{\kLinIndepAnti}{R}           
\newcommand{\skey}{\kappa}               
\newcommand{\keyseq}{\phi}               
\newcommand{\lone}[1]{\ell_1(#1)}        
\newcommand{\coefflag}{d_c}              
\newcommand{\rankflag}{\rho}             
\newcommand{\rankkey}{\mathrm{rank}}      
\newcommand{\ordk}{\omega_{\downarrow\skey}} 
\newcommand{\ordc}{\omega_{\downarrow c}}    
\newcommand{\CLG}{\mathcal{L}}           
\newcolumntype{P}[1]{>{\begin{minipage}[t]{#1}%
  \setlength{\leftskip}{0pt}\setlength{\rightskip}{0pt}%
  \setlength{\parfillskip}{0pt plus 1fil}}l<{\end{minipage}}}

\usepackage[utf8]{inputenc}
\usepackage{mathptmx}
\usepackage{etoolbox}

\makeatletter
\def\@email#1#2{%
 \endgroup
 \patchcmd{\titleblock@produce}
  {\frontmatter@RRAPformat}
  {\frontmatter@RRAPformat{\produce@RRAP{*#1\href{mailto:#2}{#2}}}\frontmatter@RRAPformat}
  {}{}
}%
\makeatother

\begin{document}


\title{All Non-Contextual Pauli Hamiltonians are Free-Fermion in Disguise}

\author{Andrew M. Projansky}
\email{andrew.m.projansky.gr@dartmouth.edu}
\thanks{A.M.P and W.W. contributed equally to this work.}
\author{Weishi Wang}
\thanks{A.M.P and W.W. contributed equally to this work.}
\author{Kevin A. Reiss}
\author{James D. Whitfield}%
\email{james.d.whitfield@dartmouth.edu}
\affiliation{%
Physics and Astronomy Department, Dartmouth College
}%

\date{\today}

\begin{abstract}
    Decomposing Hamiltonians into classically tractable fragments is central to quantum simulation. One such way of doing so is the {\it contextual subspace} decomposition, which expresses the Hamiltonian with respect to non-contextual sets of Paulis. The conditions for contextuality rely on the anticommutation and commutation of Pauli operators; it is therefore natural to make an explicit connection between non-contextuality and fermionic physics. Theoretical work is centered around two theorems: in the first, we consider non-contextual Pauli Hamiltonians and show that they are free-fermion-disguised. In the second we write these Pauli non-contextual Hamiltonians in terms of Majorana and Dirac fermionic operators, showing that they carry a Jordan--Wigner-like fermion-to-qubit encoding. These two central theorems are formalized in the proof assistant language {\it Lean}. We use this to describe the space of fermionic non-contextual Hamiltonians, and highlight their relation to the more general class of free-fermion-disguised Hamiltonians. 
    We further develop a heuristic algorithm to find
    line-graph based free-fermion fragments and their connected components of molecular Hamiltonians. We benchmark our algorithms against non-contextual searches for encoded molecular Hamiltonians and find comparable performance in identifying free-fermion solvable components. Our work has implications for what fragments to consider as useful for quantum simulation. 
\end{abstract}

\maketitle


\section{Introduction}
\label{sec:introduction}

\indent Developing quantum algorithms for Hamiltonian simulation and ground-state energy approximation often involves decomposing Hamiltonians into fragments which can be exactly or approximately solved by classical methods \cite{kitaev2002classical, Aaronson_2004, gottesmanknill}. The fragments used in these decompositions depend on both the system under study and the intended application. 
There are instances where a fragment's ground state is hard to compute, but finding a unitary which diagonalizes the fragment is efficient and sufficient for the problem \cite{patel2023extensionexactlysolvablehamiltoniansusing}. There are problems however, such as ground-state approximation for large systems, and time evolution, for which classical hardness limits applicability.

\indent The test cases for decomposition methods for ground-state energy approximation and time evolution have been centered around two-body electronic Hamiltonians and the electronic structure problem. For time evolution, methods such as double factorization and tensor hypercontraction (THC) reduce the cost of Hamiltonian simulation of molecular Hamiltonians \cite{WeibeDF, DF2, TroyerDF, necaise2026distributioncomplexityelectronicstructure, hohenstein2012thc, lee2021thc}. Quantum chemistry has a number of methods for ground-state approximation from Hartree--Fock to Density Functional Theory to post-Hartree--Fock methods \cite{Tom_MF1, Tom_MF2, szabo1996modern}. \\
\indent Here we are concerned with two families of Hamiltonian fragments: 
non-contextual Hamiltonians and free-fermion-solvable Hamiltonians. Contextuality developed as a stronger requirement than locality for hidden-variables formulations of quantum mechanics \cite{bell1966Problem, kochen1967Problem, mermin1993Hidden}. 
Non-contextual Hamiltonians have a practical use in quantum information, and can be seen as interpolating between commuting and anticommuting Hamiltonians \cite{La_Cour_Context, Ralli_ContextVQE, Kirby_ContextVQE, Kirby_Contextuality,Kirby_ContextClassic, Raussendorf_MBQC, Weaving_ContextVQE}. Free-fermion-solvable Hamiltonians have emerged from the study of Pauli Hamiltonians whose spectra are free-fermion directly, after specifying a symmetry sector, or after identifying certain \enquote{disguises} in the frustration graph of the Hamiltonian \cite{Chapman_Fermions, fermion_unified, Elman_Fermions, projansky_fermions, Tom_MF1, Tom_MF2, Ruh_2026, ruh2026graphtheoreticanalysisnongenericfreefermion}. Here, ``free fermions'' refers to systems that have no many-body interaction term.
\indent In this paper, we show that the family of free-fermion-disguised Pauli Hamiltonians contains but is not equivalent to the family of Pauli non-contextual Hamiltonians. We also show that representing the non-contextual Hamiltonians with respect to Dirac or Majorana fermion operators generalizes work into the structure of Pauli non-contextual Hamiltonians. 
This work builds off of insights from two works: Patel and Izmaylov \cite{patel2024exactly}, and Patel, Yen, and Izmaylov \cite{patel2023extensionexactlysolvablehamiltoniansusing}. In those papers, Patel leads the discussion on how \enquote{exactly solvable} Hamiltonians can emerge from considering extensions of Hamiltonians via a set of Abelian symmetries of the Hamiltonian. In these papers, exactly solvable refers to Hamiltonians which can be diagonalized efficiently. Their work builds on literature about the sub-Lie algebras of the Pauli-group \cite{aguilar2024classificationpauliliealgebras}. 

Our theoretical discussion centers around the literature of \enquote{free-fermion solvability,} a subfield which studies Hamiltonians by asking if they admit a mapping to a free or free but \enquote{disguised} Hamiltonian \cite{Chapman_Fermions, fermion_unified, Elman_Fermions}. In doing so, we unify the literature of non-contextuality and free-fermion solvability. We extend the theoretical work of Patel et al.\ by considering Majorana or Pauli operator independence, which allows us to write variants of non-contextual Dirac fermion Hamiltonians. We also extend their theoretical work by considering the effect of Kitaev-honeycomb-like cycle symmetries which can still be present, even when the frustration graph is a line graph \cite{Kitaev_2006, Chapman_Fermions}. We reproduce and extend the numerical calculations of Patel et al. by presenting a new graph-search algorithm and show that it performs better than the original numerics at identifying line-graph components for encoded molecular Hamiltonians. Most vital to the outlook of the paper, we review and discuss the complexity-theoretic constraints which restrict the practical applications in most use cases of non-contextual Hamiltonians. 
For practical applications, the line graph description of free-fermion Hamiltonians is a fundamental fragment primitive with utility in quantum simulation. 

\indent This paper is structured as follows. In Sec.~\ref{sec:preliminaries} we review preliminaries about fermions, fermion-to-qubit encodings, and graph theory. In Sec.~\ref{sec:Background}, we introduce contextuality, free-fermion solvability, and free-fermion disguises. There, we prove that all non-contextual Pauli Hamiltonians are twin-free-fermion-disguised. In Sec.~\ref{sec:fermionic-encodings} we show the duality between non-contextual Pauli Hamiltonians and Jordan--Wigner-like fermion-to-qubit encodings, and introduce fermionic non-contextual Hamiltonians. In Sec.~\ref{sec:molecular-hamiltonians} we investigate non-contextual and free-fermion decompositions for molecular Hamiltonians and present numerical results from our line-graph-based search algorithm, compared with non-contextuality heuristics. Finally, we conclude in Sec.~\ref{sec:conclusion} with discussion of open questions, including extensions to free fermions in disguise and parafermions.

\begin{table*}[t]
    \centering
    \caption{Table summarizing the Hamiltonians considered in this work. Summarized are
      the complexity class of finding their ground‑state energies, and
      the gate cost of implementing Hamiltonian simulation for an $n$-qubit Hamiltonian of the particular kind.  The unitaries
    which diagonalize $\noncontextualHamiltonian$ and $\freeFermionHamiltonian$
      can be solved for efficiently \cite{patel2023extensionexactlysolvablehamiltoniansusing}. However, 
      in the worst case, it may still be exponentially
      costly in the number of qubits 
      to do simulation without Trotter error $\epsilon$ (defined in Appendix~\ref{app:qcomp}). }
    \label{fig:hamiltonian-table}

    \begin{tabular}{l c c c c}
        \toprule
        \textbf{Hamiltonian} &
         &
         &
        \textbf{\(E_{\min}\)} &
        \textbf{Hamiltonian Simulation for fixed small $t$} \\
        \midrule
        Stabilizer & $\stabilizerHamiltonian$ & Eq.~\eqref{eq:stabilizer-fermionic-hamiltonian} & $\mathrm{P}$ & $\operatorname{poly}(n)$ \\
        Commuting (Poly) & $\commutingHamiltonian$ & Eq.~\eqref{eq:commuting-fermionic-hamiltonian} & $\mathrm{NP}$ & $\operatorname{poly}(n)$ \\
        Anticommuting & $\anticommutingHamiltonian$ & Eq.~\eqref{eq:anticommuting-hamiltonian} & $\mathrm{P}$ & $\operatorname{poly}(n)$ \\
        Free-fermion (line graph) & $\lineGraphFreeFermionHamiltonian$ & Eq.~\eqref{eq:line-graph-free-fermion-hamiltonian} & $\mathrm{P}$ & $\operatorname{poly}(n)$ \\
        Interacting fermion & $H_{\mathrm{I}}$ & Eq.~\eqref{eq:interacting-fermionic-hamiltonian} & $\mathrm{QMA}$ & $\operatorname{poly}(n,1/\epsilon)$  \\
        \midrule
        Non-contextual & $\noncontextualHamiltonian$ & Eq.~\eqref{eq:noncontextual-hamiltonian} & $\mathrm{NP}$ & $\operatorname{poly}(n,1/\epsilon)$ \\
        Twin-free-fermion-disguised & $\freeFermionHamiltonian$ & Eq.~\eqref{eq:twin-vertex-disguised-hamiltonian} & $\mathrm{NP}$ & $\operatorname{poly}(n,1/\epsilon)$ \\
        \bottomrule
    \end{tabular}

\end{table*}

\begin{figure*}
    \centering
    \includegraphics[width=0.7\textwidth]{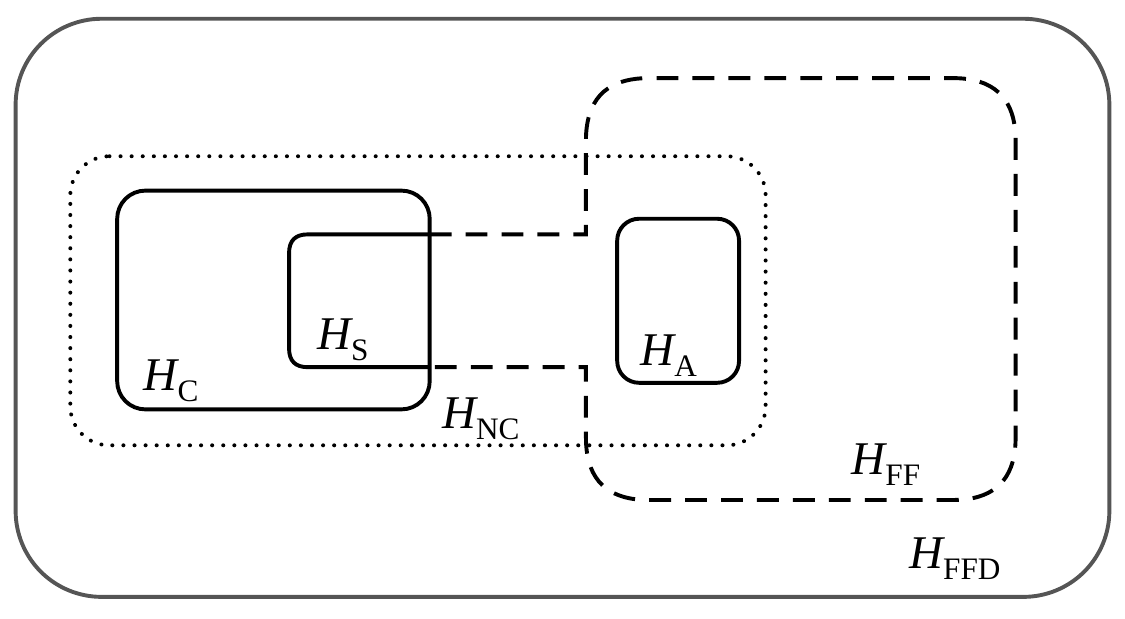}
    \caption{The inclusion chart for the Hamiltonians in Table~\ref{fig:hamiltonian-table}, and throughout this paper, except for the interacting fermionic Hamiltonian.}
    \label{fig:inclusion}
\end{figure*}

\section{Preliminaries}
\label{sec:preliminaries}

\subsection{Fermions and Fermionic Hamiltonians}
\label{sec:fermions-and-fermionic-hamiltonians}

\indent Molecular Hamiltonians, and any quantum system of electrons, are often studied by discretization with a choice of single-particle basis functions (i.e., orbitals). These are then reformulated in second quantization as linear combinations of operator strings (i.e., operator products). This discretization induces a finite-dimensional fermionic 
Fock space of dimension $2^m$, where each basis state represents the fermionic occupancy in $m$ quantized modes. The canonical raising and lowering operators on Fock space are the  Dirac fermion operators, $a_p^{\dagger}$ and $a_p$. Let the commutator and the anticommutator be defined as  $\comm{A}{B} = AB-BA$ and  $\ac{A}{B} = AB+BA$. Then the canonical anticommutation relations are $\ac{a_p}{a_q} = \ac{a_p^{\dagger}}{a_q^{\dagger}} = 0$, and $\ac{a_p^{\dagger}}{a_q} = \delta_{pq}$. \\ 
\indent For studies that utilize this anticommutation structure, 
we define the Majorana fermion operators as: 
$\gamma_{{2p-1}} = (a_p+a_p^{\dagger})$ and $\gamma_{{2p}} = \mathrm{i}({a_p^{\dagger} - a_p})$. The Majorana operators obey the anticommutation relation, 
\begin{equation}
    \ac{\gamma_p}{\gamma_q} = 2\delta_{pq}.
    \label{eq:majorana-anticommutation}
\end{equation}
\indent All Majorana operators pairwise anticommute and square to the identity. To represent these fermionic operators, we identify the Fock space as a $2^m$ dimensional Hilbert space, for $m$ fermionic modes. 
An operator basis for all fermionic operators can be expressed using products of the $2m$ Majorana operators, e.g., $O_b = \gamma_1^{b_1} \gamma_2^{b_2}\gamma_3^{b_3} ... \gamma_{2m}^{{b_{2m}}}$ for a $2m$-bit bitstring $b$. The total number of operators that can be expressed in this form is $2^{2m} = 4^m$. This is the number of Hermitian matrices needed for a basis of the $2^m \times 2^m$ Hermitian matrices. \\
\indent In this paper, when we discuss Hamiltonians, we are interested in three key aspects of them: if solving the ground-state energy problem for a Hamiltonian is in one complexity class or another, if its corresponding Hamiltonian simulation has a known implementation in polynomial depth with no Trotter error, and if a Hamiltonian contains the Hamiltonians of another family of problems. For the Hamiltonians considered throughout this paper, a table reviewing the complexity of approximating ground-state energy and the Trotter simulation costs can be seen in Table~\ref{fig:hamiltonian-table}. A chart diagramming inclusion of Hamiltonians in other classes can be seen in FIG.~\ref{fig:inclusion}. \\
\indent Hermitian fermionic Hamiltonians can be faithfully represented by both Dirac  operators and Majorana fermion operators. 

Fermionic Hamiltonians 
can be classified as interacting or non-interacting. Non-interacting fermionic Hamiltonians 
can be expressed as 
\begin{align}
    \lineGraphFreeFermionHamiltonian = {\mathrm{i}}\sum_{p,q} h_{pq}\gamma_p \gamma_q,
    \label{eq:line-graph-free-fermion-hamiltonian}
\end{align}
with $h_{pq}$ real and antisymmetric. Such Hamiltonians are composed solely of quadratic products of Dirac or Majorana fermion operators. All Hamiltonians of this kind can be efficiently solved classically with a Bogoliubov transform, rotating the 
space of fermionic modes into a diagonalizing single-particle mode basis \cite{Bogolyubov:1947zz, Surace_2022}. Interacting fermionic Hamiltonians are expressed generally, with respect to Majorana operators as 
\begin{align}
    H_{\mathrm{I}} ={}& {\mathrm{i}}\sum_{p,q}  h_{pq} \gamma_p \gamma_q + \sum_{p,q,r,s}  V^{(2)}_{pqrs} \gamma_p \gamma_q \gamma_r \gamma_s
    \label{eq:interacting-fermionic-hamiltonian}
\end{align}

For this work, there are two other relevant classes of fermionic Hamiltonians. 
The first is pairwise anticommuting fermionic Hamiltonians. They are written with respect to the Majorana operators as
\begin{equation}
    \anticommutingHamiltonian = \sum_r \coeff_r \gamma_r,
    \label{eq:anticommuting-hamiltonian}
\end{equation}
where $\coeff_r$ is the real coefficient associated with $\gamma_r$.
Pairwise anticommuting Hamiltonians can be efficiently solved classically because they generate a compact Lie algebra which can be mapped to the algebra of line-graph-free-fermion Hamiltonians \cite{Chapman_Fermions}. 

The second relevant class is commuting fermionic Hamiltonians. Such Hamiltonians are related to {\it Majorana stabilizer groups} \cite{Bravyi_MajoranaCodes, Mudassar_2024}. Consider a set of $2m$ Majorana operators $\{\gamma_r\}$, and consider a pairing of all Majorana operators as $\{\gamma_1 \gamma_2, \gamma_3 \gamma_4, ..., \gamma_{2m-1}\gamma_{2m} \}$. The elements of this set are also 
a set of $m$ independent and pairwise commuting operators. An independent set of Paulis is one such that no element can be expressed as the product of other elements of the set. This set defines a stabilizer group, and generates a set of $2^m$ commuting fermionic operators. Letting the elements of the matching be labeled $\Delta_r$, we can express 
stabilizer fermionic Hamiltonians as 
\begin{equation}
    \stabilizerHamiltonian = \sum_r \coeff_r \Delta_r
    \label{eq:stabilizer-fermionic-hamiltonian}
\end{equation}
and general commuting Hamiltonians as 
\begin{equation}
    \commutingHamiltonian = \sum_r \coeff_r \Delta_r + \sum_{p,q} \coeff_{pq} \Delta_p \Delta_q + \cdots .
    \label{eq:commuting-fermionic-hamiltonian}
\end{equation}
for real coefficients $f$. 


\subsection{Fermion-to-Qubit Encodings and Qubit Hamiltonians}
\label{sec:fermion-to-qubit-encodings}
\indent To simulate fermions on a quantum computer, we need to pick some representation of the anticommuting operators. The matrix representation for anticommuting operators can be written with respect to the Pauli basis. Let $\mathbf{P} = \langle \mathrm{i}\Id, X_1, Z_1, ..., X_n, Z_n\rangle$ be the generators of the Pauli group on $n$ qubits, the single-qubit $X_j, Z_j \in \mathbf{P}$ operators and phase factors $\{\pm 1, \pm \mathrm{i} \}$. All Paulis (i.e. Pauli operators) can be generated from this set, recalling $X_p Z_p =\mathrm{-i}Y_p$. \\

\indent By attempting to represent the Dirac or Majorana fermion operators on the fermionic Fock space, one can derive a fermion-to-qubit transformation by realizing that the $m$-mode Fock space is composed of the same basis states as an $n\!=\!m$ qubit Hilbert space with respect to the $Z$ basis. This canonical fermion-to-qubit transformation is the Jordan--Wigner transform, 
\begin{equation}
    \gamma_{2p-1} = \left[\prod_{q=1}^{p-1}(Z_q)\right]X_p \hspace{1.3cm} \gamma_{2p} = \left[\prod_{q=1}^{p-1}(Z_q)\right]Y_p.
    \label{eq:jordan-wigner-transform}
\end{equation}

Jordan--Wigner encodes all bitstrings from Fock space into the equivalent bitstrings in the Hilbert space of the quantum computer, and maps all fermionic operators to qubit operators with the same matrix representations. In the Jordan--Wigner encoding, number operators are local, but this mapping comes at the cost of requiring arbitrary Pauli operators which may be highly non-local. \\
\indent The Jordan--Wigner Majorana operators are a set of $2n$ pairwise anticommuting Paulis over $n$ qubits. Conjugating this set by {\it any} Clifford unitary results in a set of $2n$ pairwise anticommuting Paulis, and constitutes a valid fermion-to-qubit encoding \cite{gottesmanknill, Aaronson_2004, chiew2026optimalfermionqubitmappingsquadratic, Yu_2025}. This observation leads to a multitude of different fermion-to-qubit encodings (i.e., the number of qubits equals the number of modes) with different properties, such as the parity encoding, Bravyi--Kitaev encoding, ternary encoding, balanced Jordan--Wigner, and Sierpinski \cite{harrison2024sierpinskitrianglefermiontoqubittransform, chiew2026optimalfermionqubitmappingsquadratic, Jiang_2020, Miller_2023}. \\
\indent There also exist more general fermion-to-qubit encodings, sometimes called one-to-many or lattice encodings, in which the number of modes between fermions and qubits changes \cite{ Derby_2021, Setia_2019}. { All encodings made up of Paulis (sometimes described as Jordan--Wigner like) can be understood 
as sets of pairwise anticommuting Paulis. 
\indent In work by Sarkar and (van den) Berg \cite{sarkar2019setscommutinganticommutingpaulis}, they prove a number of 
facts about sets of pairwise anticommuting Paulis which we review here: 
\begin{itemize}
    \item The maximum number of anticommuting Paulis on $n$ qubits is $2n+1$. We will refer to sets which saturate this bound as maximal. 
    \item Non-maximal sets of anticommuting Paulis can still be \enquote{complete}, in that no more anticommuting terms can be added. 
    \item Complete encodings are such that the product of all the Paulis 
    is proportional to the identity, and the set of operators is of odd cardinality.  
    \item Given a non-maximal incomplete set over $n$ qubits, there exist a randomized algorithm to extend that set to $2n+1$ anticommuting elements. 
\end{itemize}
We will utilize the implications of these facts throughout this paper. 
Two 
immediate implications are as follows:
\begin{enumerate}
\item Complete sets of anticommuting Paulis are not independent. Let $Q = \gamma_1 \gamma_2\cdots \gamma_{2n+1}$ be the product of a maximal set of anticommuting Paulis on $n$ qubits, then $Q
~{\propto}~ \Id$. This, combined with the fact that each $\gamma_p$ is a Pauli which squares to $\Id$, implies that given $2n$ Majorana operators $\{\gamma_i\}$, the final Majorana operator which completes the encoding is $\gamma_{2n+1} \propto \gamma_1 \gamma_2 ... \gamma_{2n}$. Throughout this paper, we will refer to the non-independent Majorana operator in a complete set as the {\it parity} operator of the encoding. For the Jordan--Wigner encoding, the natural parity operator is $Z_1Z_2...Z_n$. We have the freedom however, to swap any Majorana operator in our incomplete set with the parity operator, and switch the definition of which Majorana operator is independent with respect to the others. 
    \item Given a one-to-many encoding, there exist multiple extensions to complete encodings. Particularly, there exist multiple Clifford transformations which conjugate a subset of the Jordan--Wigner Majorana operators on the full physical space of qubits to the desired Majorana operators of the one-to-many encoding.
\end{enumerate}
Consider as an example the $[[5,1,3]]$ code \cite{laflamme1996perfect, bennett1996mixed, eczoo_stab_5_1_3}, which contains logical operators $\tilde X = X_1X_2X_3X_4X_5$ and $\tilde Y = Y_1Y_2Y_3Y_4Y_5$, and the stabilizers $\{X_1Z_2Z_3X_4,\,X_2Z_3Z_4X_5,\,X_1X_3Z_4Z_5,\,Z_1X_2X_4Z_5\}$. The operators $\tilde X$ and $\tilde Y$ form a non-maximal 
set of anticommuting operators over $5$ qubits. This means we can extend this set to a set of $11 = 2\times 5 + 1$ anticommuting Paulis. One possible extension is 
\begin{align}
    \{& X_1X_2X_3X_4X_5, Y_1Y_2Y_3Y_4Y_5, Y_1Z_2Y_3Y_4Y_5, Z_1X_3Z_4Y_5, \notag\\
    & -Y_1X_2X_3Y_4Z_5, Y_1Y_3X_4Y_5, -Z_1X_2Z_3X_4Z_5, Y_1X_2Y_3Y_4, \notag\\
    & -Z_1Z_3Z_4, X_2Z_3Y_4Z_5, -Y_3Z_4Y_5\}.
    \label{eq:five-qubit-extension}
\end{align}
This is the extension such that $-\mathrm{i}\gamma_{3}\gamma_4 = X_1Z_2Z_3X_4$, $-\mathrm{i}\gamma_{5}\gamma_6 = X_2Z_3Z_4X_5$, and so on, with $\gamma_{11}$ being a parity operator. From the extension of the encoding we can find anticommuting elements which form the stabilizers of the code. Furthermore, we can find the Clifford transformation which maps the Jordan--Wigner encoding to the encoding which defines the $[[5,1,3]]$ code, which also serves as the encoding Clifford for the code. 

}


Fermion-to-qubit encodings provide a tool for translating between qubit and 
fermionic Hamiltonians. For the fermionic Hamiltonians previously considered, there are Pauli 
equivalents by replacing Majorana operators with Paulis. The expressions below are re-expressions of Eqs.~\eqref{eq:line-graph-free-fermion-hamiltonian}--\eqref{eq:commuting-fermionic-hamiltonian}, and are therefore left unnumbered. 

\subsection{Frustration Graphs and Line Graphs}

\indent Graphical representations serve as a convenient tool for finding properties to exploit in a Hamiltonian to make time evolution or ground-state energy approximations easier. One such representation is the frustration graph of a Hamiltonian. 

\begin{identity}[Frustration Graph]
    Let $H$ be a Hamiltonian written with respect to a basis of operators $\{ O_p\}$ where every two distinct operators either commute or anticommute. The frustration graph $\mathcal{G}$ is defined by associating each element $O_p$ with a vertex 
    $v_p$ in the vertex set $V$, and defining the edge set
    $E$ as $(v_p,v_q)\in E$ if and only if $[O_p, O_q]_+ = 0$.
\end{identity}

\indent In other words, the edges in a frustration graph connect operators that anticommute. For example, the frustration graph for a commuting or stabilizer Hamiltonian ($\commutingHamiltonian$ and $\stabilizerHamiltonian$ respectively) is a set of disconnected vertices. The frustration graph for an anticommuting Hamiltonian $\anticommutingHamiltonian$ is a clique. \\
\indent The frustration graph for a free-fermion Hamiltonian $\lineGraphFreeFermionHamiltonian$ has been shown to be a line graph \cite{beineke1968derived}. Line graphs are a special family of intersection graphs, originally studied by Erd\"os, that represent graphs based on the intersections of subsets of a set of elements \cite{Erdos1966Representation}. Line graphs are closely related to Johnson graphs, and we give the line graph definition in terms of Johnson graph \cite{Holton_Sheehan_1993, harary1969graph}. 

\begin{identity}[Johnson Graphs and Line Graphs]
    The Johnson graphs $J(n,s)$ are defined with a vertex set of $s$-element subsets of a set of $n$ elements, and edges if the two vertices (subsets) contain $s-1$ intersections. Line graphs are the induced subgraphs of $J(n,2)$, where an induced subgraph is a graph constructed by removing some subset of points and all edges connected to those points. 
\end{identity}

Line graphs are well studied in large part because determining if a graph is a line graph is efficient \cite{ROUSSOPOULOS1973108}.


\section{Non-Contextual Pauli and Free-Fermion-Solvable Hamiltonians}
\label{sec:Background}

\subsection{Contextuality}
\label{sec:contextuality}

\indent The question of whether observable quantities have certain, pre-defined values regardless of our observation of them was central to the study of quantum theory throughout its early history \cite{einstein1935Can}. 
To explain which eigenvalue is realized when a quantum state exhibits non-zero dispersion for a given observable, attempts were made to construct a ``hidden-variables'' theory which tracks additional parameters beyond those in the wavefunction \cite{debroglie1930Introduction, bohm1952Suggested, bohm1952Suggesteda, mermin1993Hidden}. 
After an initial refutation of the possibility of hidden-variables theories by von Neumann \cite{vonneumann2018Mathematical, albertson1961Neumanns}, 
J.S. Bell laid out two no-go theorems that specify which conditions are incompatible with a hidden-variables description of quantum theory. The first \cite{bellt}
(in publication order)
forbids \textit{local} hidden-variables theories.  The second \cite{bell1966Problem}, derived independently by Kochen and Specker \cite{kochen1967Problem} (and therefore referred to as the Bell--Kochen--Specker theorem) forbids non-\textit{contextual} hidden-variables theories.

\begin{figure}
    \centering
    \includegraphics[width=\linewidth]{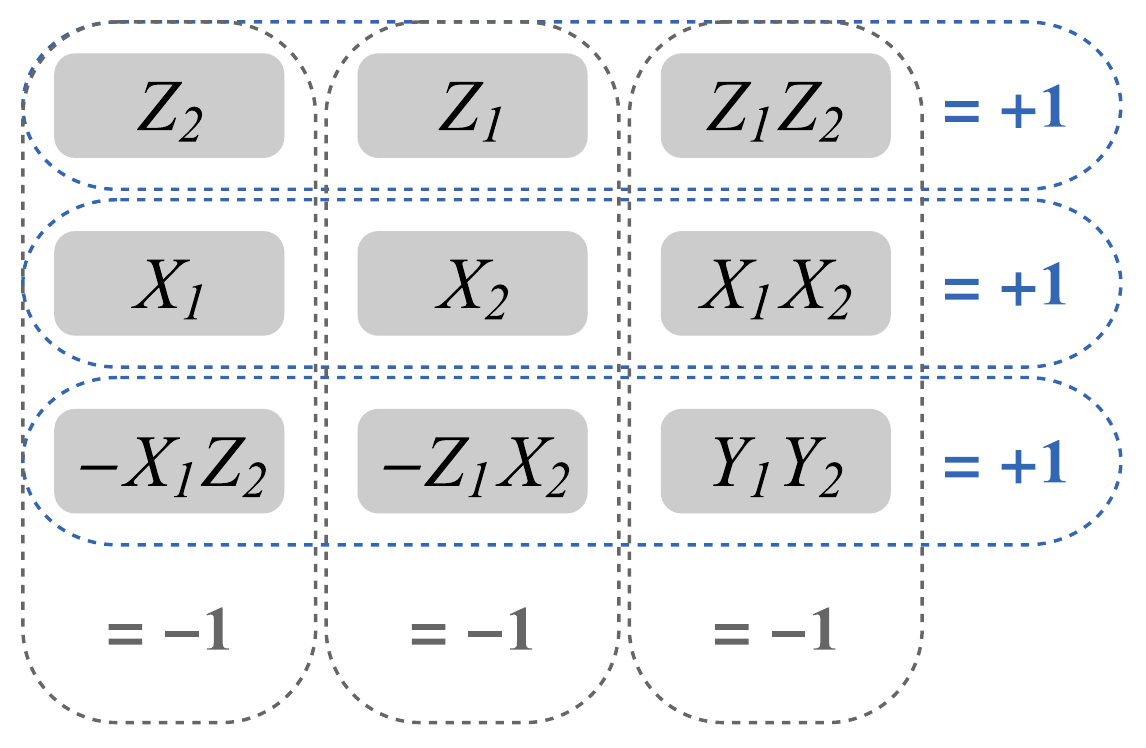}
    \caption{The Peres-Mermin square, the canonical example of a contextual set of operators on a four-dimensional Hilbert space. A measurement on the set of operators in each row must multiply to $+1$, while the columns multiply to $-1$. No assignment of $\pm1$ to each operator can simultaneously satisfy all six constraints.}
    \label{fig:PMSquare}
\end{figure}

Contextuality arises when observables cannot consistently be given values pre-measurement. The set of observables may form pairwise commuting subsets which obey restrictions that are incompatible with the restrictions imposed on other subsets. A canonical example is the Peres-Mermin square \cite{PERES1990107, mermin1990Simple,mermin1993Hidden}, a construction of which is shown in FIG.~\ref{fig:PMSquare}. The $3\times 3$ grid of Pauli operators is arranged such that the elements of each row commute with one another, and similarly for columns. Therefore, any row or column specify a ``context'' of three operators which can be simultaneously observed. However, each row multiplies to $\Id$, while each column multiplies to $-\Id$. The product of measurements on any complete row must therefore be $+1$, and similarly column measurements must multiply to $-1$. However, the row constraints are inconsistent with the column constraints, since the rows require an even number of $-1$ values while columns require an odd number. Therefore no $3\times 3$ grid can be pre-filled with eigenvalues $+1$ or $-1$ to satisfy these constraints simultaneously. Thus, no assignment of context-independent values $+1$ or $-1$ to the nine observables can satisfy the product constraints for every row and column. The Peres-Mermin square therefore demonstrates the impossibility of a non-contextual value assignment consistent with the quantum predictions. In a hidden-variables theory that assigns definite values, those values must depend on the measurement context.

Contextuality indicates that the value received from measuring an observable must inherently depend on the full set of observables with which it was measured. 
For a hidden-variables theory to specify all possible values of observables, it would need to enumerate different sets of values for all possible measurements (\enquote{dispositions of the apparatus} \cite{bell1966Problem}). Since its philosophical development, contextuality has evolved from describing a necessary feature of quantum theories to a feature of sets of observables for the purpose of identifying non-classical properties. In \cite{Kirby_Contextuality} it was shown that for a contextual set of Pauli operators, there exists some Pauli $A$  such that both $A$ and $-A$ can be determined (produced by multiplying elements of the set together). This extends the concept of incompatible observables beyond non-commutation. \\
\indent We can now introduce contextuality and the lack of it in more practical terms.

 \indent  
\begin{identity}[Non-Contextual Operators]
\label{id:nc}
    Consider a set of operators $S = \{O_r\}$ which pairwise either commute or anticommute. Let $G \subseteq S$ be the set of all elements which pairwise commute with all other elements in $S$, and denote $S' = S \setminus G$. $S$ is non-contextual if and only if commutation is an equivalence relation on $S'$.
\end{identity}

Said otherwise, $S$ is non-contextual if and only if for any $A, B, C \in S'$, $[A,B] = 0$ and $[A,C] = 0$ implies $[B, C] = 0$. Recalling FIG. \ref{fig:PMSquare}, the center element $X_2$ commutes with every other operator in the second row and the second column. Although $X_2$ commutes with, e.g., $Z_1$ and $X_1$, the latter two operators do not commute with each other (as they do not share a row or column). 

All non-contextual Pauli Hamiltonians can be generated under the anticommutator by a pairwise anticommuting set of Paulis $\mathcal{A}$ and a set of Paulis $G$ which commute with each other 
    \cite{Ralli_ContextPauli, patel2023extensionexactlysolvablehamiltoniansusing}. Specifically, all terms are generated from elements and products of elements from $\mathcal{A}\cup G$. \\
\indent Let each anticommuting element of $\mathcal{A}$ represent a Majorana operator $\gamma_p$ (via a fermionic encoding), and  let $L=|\mathcal{A}|\leq {2n+1}$ be the number of pairwise anticommuting operators. Here $L<2n+1$ for non-maximal encodings. The set of commuting Pauli strings is $G$ and is multiplicatively generated by a set of $\kLinIndepComm$ linearly independent commuting Pauli operators $\{C_j\}$ and $|G|=2^\kLinIndepComm$. Accordingly, a non-contextual Hamiltonian can have the following compact expression:
\begin{equation}
    \noncontextualHamiltonian = \sum_{r=1}^{L} \gamma_r p_r (C_1, C_2, ..., C_{\kLinIndepComm}) + p_C(C_1, C_2, ..., C_{\kLinIndepComm})
    \label{eq:noncontextual-hamiltonian}
\end{equation}
where $p_r(C_1, C_2, ..., C_{k})$ and $ p_C(C_1, C_2, ..., C_{\kLinIndepComm})$ are polynomials over Pauli operators \cite{patel2023extensionexactlysolvablehamiltoniansusing}. We define these polynomials such that $p(x_1,x_2,...x_k) =\sum_b^{2^k} c_b x_1^{b_1} x_2 ^{b_2} ... x_{\kLinIndepComm} ^{b_{\kLinIndepComm}}$ where the summation is over bitstrings $b$ each of length $\kLinIndepComm$ and the coefficients $c_b$ change with each polynomial.\\ 
\indent The number of terms in non-contextual Hamiltonians depends on the number of anticommuting operators, relative to the total number of qubits those operators act on. As stated earlier, the maximum size of a set of pairwise anticommuting operators over $n$ qubits is at most $2n+1$. 
For odd values of $L$, we have $k = \frac{2n+1-L}{2}$. 



\subsection{Free-Fermion Solvability}
\label{sec:free-fermion-solvability}

\indent Recent years have seen the development of methods for characterizing free-fermion symmetries in spin Hamiltonians via graphical invariants in the frustration graphs of the Hamiltonians \cite{Chapman_Fermions, fermion_unified, Elman_Fermions}. The most basic class of free-fermion-solvable systems are those solvable via their mapping to a line graph. 

\begin{theorem}
    [Chapman and Flammia, 2020, Existence of Free-Fermion Solution \cite{Chapman_Fermions}]
    For an $n$-qubit Hamiltonian $H$, written with respect to the Pauli basis, a Jordan--Wigner-like fermion-to-qubit encoding exists such that the qubit Hamiltonian represents a free-fermion Hamiltonian (with possible non-trivial cycle symmetries) if the frustration graph $\mathcal{G}$ of $H$ is the line graph of another graph $\mathcal{R}$ (i.e., $\mathcal{R}$ is the root graph of $\mathcal{G}$).
\end{theorem}

\begin{figure}[htbp]
    \centering

    \subfloat[\label{fig:fg_h1s}]{
        \includegraphics[width=0.8\columnwidth]{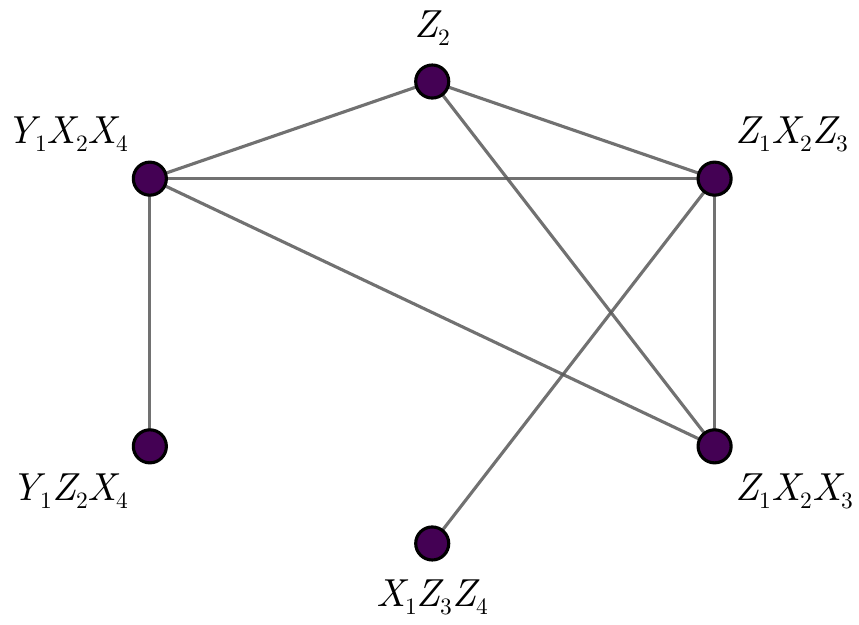}
    }
    

    \subfloat[\label{fig:fg_h2s}]{
        \includegraphics[width=0.8\columnwidth]{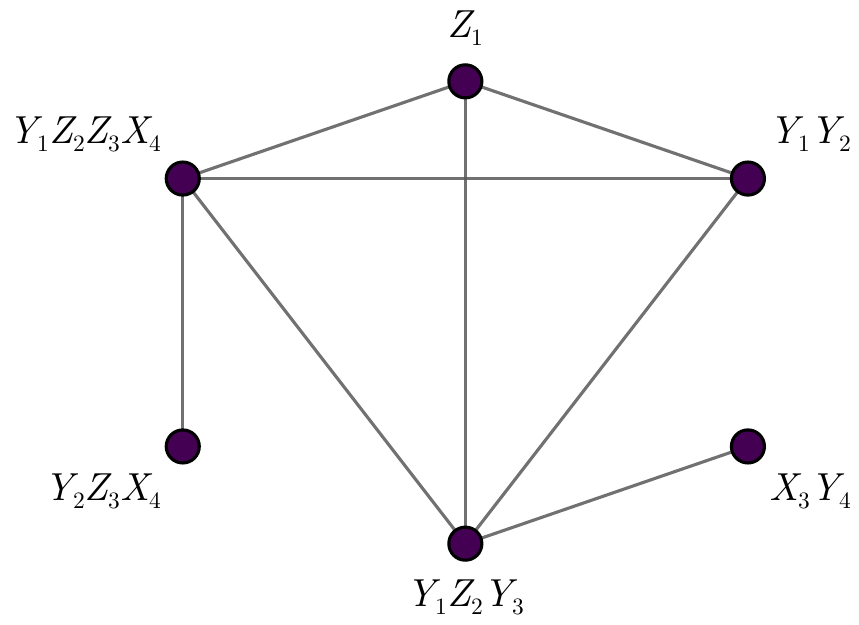}
    }

    \caption{The frustration graphs of $H_S^1$ and $H_S^2$, in (a) and (b), respectively. Though laid out differently, the two graphs are isomorphic, with a vertex-to-vertex mapping defined by pairing terms in the ordered sums $H_S^1$ and $H_S^2$.}
    \label{fig:fg_h1sh2s}
\end{figure}

For this reason, we are particularly interested in the connected components of a frustration graph that are line graphs, i.e., connected line-graph components. Line-graph spin Hamiltonians are generalizations of Jordan--Wigner-solvable spin models. For example, consider the following two spin Hamiltonians: 
\begin{align}
    H_S^1 = & Z_2 - Z_1 X_2 X_3 + Y_1 X_2 X_4  + \notag \\
    & Y_1 Z_2 X_4 + X_1 Z_3 Z_4 - Z_1 X_2 Z_3,
    \label{eq:line-graph-spin-hamiltonian1} \\ 
    H_S^2 = & Z_1 + Y_1 Y_2 + Y_1 Z_2 Z_3 X_4 +  \notag \\
    & Y_2 Z_3 X_4 + X_3 Y_4 + Y_1 Z_2 Y_3.
    \label{eq:line-graph-spin-hamiltonian2}
\end{align}
\indent Their frustrations graphs are shown in FIG.~\ref{fig:fg_h1sh2s}. These two Hamiltonians are neither commuting nor anticommuting, and are not non-contextual. Their frustration graphs are both line graphs and isomorphic to each other. Thus, they can be generated by single Majorana operators and quadratic products with respect to different choices of fermion-to-qubit encoding. In this case, the first Hamiltonian comes from the ternary tree encoding \cite{Jiang_2020}, while the second comes from the Jordan--Wigner encoding \cite{projansky_fermions, chiew2026optimalfermionqubitmappingsquadratic}. Both can be mapped to the fermionic Hamiltonian 
\begin{align}
    H_F = & \, \mathrm{i}\left( -\gamma_1\gamma_2 + \gamma_1\gamma_4 + \gamma_1\gamma_7
          + \gamma_3\gamma_7 - \gamma_6\gamma_8 + \gamma_1\gamma_6 \right).
    \label{eq:line-graph-fermionic-hamiltonian}
\end{align}

\indent There are many instances of Hamiltonians which are not line graphs, but still retain some free-fermion structure. This relates to the so-called {\it free fermions in disguise} \cite{Chapman_Fermions, fermion_unified, Elman_Fermions}. There are various kinds of disguises free fermions can take: the most basic, and the ones central to this paper, are indicated by twin-vertex symmetries in frustration graphs.

\begin{figure}
    \centering
    \includegraphics[width=0.8\linewidth]{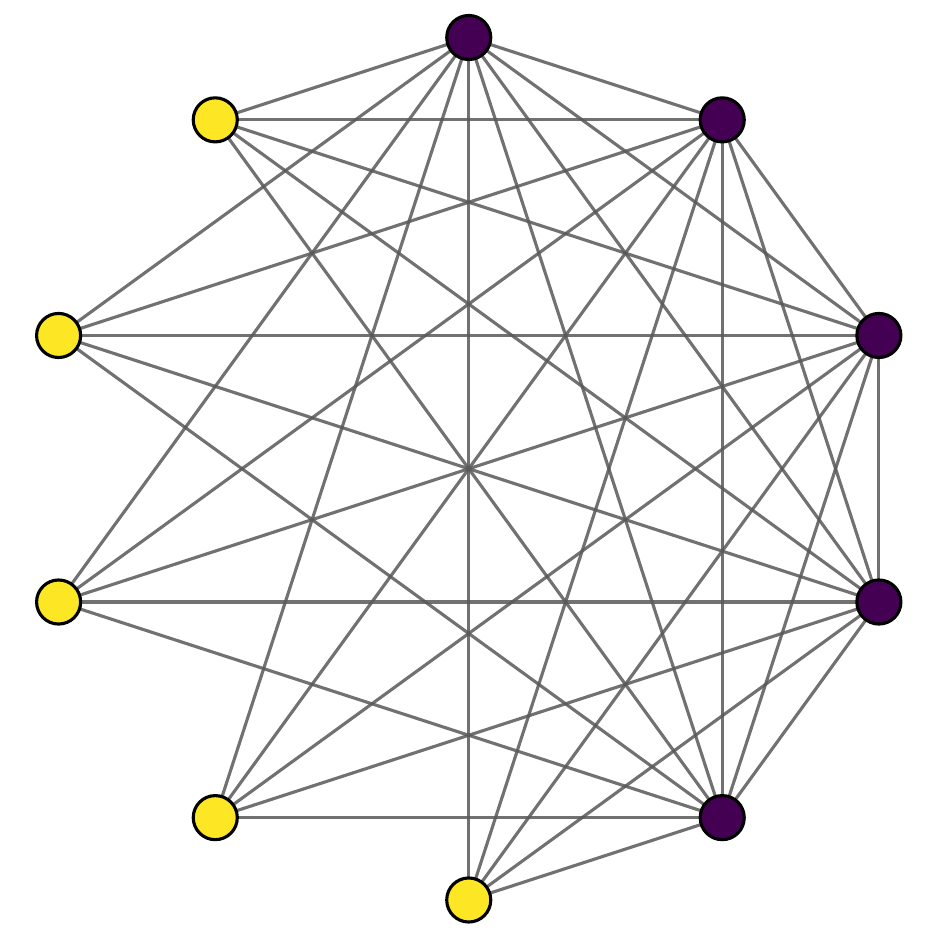}
    \caption{An example of twin vertices in a simple graph. All five purple vertices are twins (among themselves) with respect to all five yellow vertices, and vice versa. In other words, twin vertices share the same neighborhoods (excluding themselves).}
    \label{fig:twins}
\end{figure}

\begin{identity}[Twin Vertices]\label{id:twins}
    Given a graph $G$ 
    with vertex set $V$ and edge set $E$, vertices $u$, $v \in V$ are twin vertices if, for every other 
vertex $w \in V$, $(u,w) \in E$ if and only if $(v,w) \in E$.
\end{identity}
Said otherwise, twin vertices share the same set of incident edges. An example of twin vertices can be seen in FIG.~\ref{fig:twins}. Twin vertices, and products of the Pauli terms on twin vertices, generate conserved charges which can be used to block diagonalize the Hamiltonian. Disguised free-fermion models, including ones with twin vertices, are models such that in each block-diagonal section the Hamiltonian is line-graph free-fermion solvable.

It is also important to mention {\it cycle symmetries}. Cycle symmetries of line graphs are Abelian Pauli symmetries of the Hamiltonian. 

\begin{identity}[Cycle symmetries]
Let the frustration graph $\mathcal{G}$ of $H\!=\!\sum_r \coeff_r O_r$ be the line graph of another root graph $\mathcal{R}$, with each Pauli term $O_r$ representing an edge of $\mathcal{R}$. For any cycle $\mathsf{c}$ in $\mathcal{R}$, i.e., a closed loop of $\ell$ edges,
\begin{equation}
    C_{\mathsf{c}}=\mathrm{i}^{-\ell}\prod_{r\in\mathsf{c}}O_r,
\end{equation}
is the product ordered along the path taken in the cycle, i.e., a term cycle. Subsequently, the term cycles obtainable from $H$ satisfy
\begin{equation}
    [C_{\mathsf{c}},O_r]=0, \quad
    [C_{\mathsf{c}},C_{\mathsf{c}'}]=0, \qquad 
    \forall \mathsf{c},\mathsf{c}', r.
\end{equation}
These term cycles generate an Abelian group of cycle symmetries
of $H$.
\end{identity}

Critical to our analysis, {\it detecting a line graph is not enough to guarantee solvability, but it implies solvability when specifying a sector of the Abelian symmetry group.} This kind of symmetry is the symmetry present in the Kitaev honeycomb model: the honeycomb graph is a line graph, but is only solvable by a \enquote{reduction to free fermions in a static $\mathcal{Z}_2$ gauge field} \cite{Kitaev_2006}. For the numerics we test in this paper to compare faithfully with the numerics of Patel and Izmaylov \cite{patel2023extensionexactlysolvablehamiltoniansusing, patel2024exactly}, we do not specify whether we detect line graphs with cycle symmetries. In general, cycle symmetries introduce classical overhead in the same way twin vertex symmetries do. The numerical examples we consider are also of small enough size that the computational restrictions that arise from cycle symmetries do not affect performance. Due to the efficient detection of line graphs and their Abelian cycle symmetries, it is practical to search for line graphs with no or a small number of cycle symmetries.  \\ 
\indent The Abelian symmetries in the free-fermion literature are the same as the commuting symmetries of non-contextual Hamiltonians: both are symmetries of the anticommuting terms $\{\gamma_p\}$, and thus by definition quadratic products of anticommuting operators $\{\gamma_p \gamma_q\}$.

\begin{align}
    \freeFermionHamiltonian = &\sum_r \gamma_r p_r (C_1, C_2, ..., C_{k}) + \notag\\
    & \sum_{r<s}\gamma_r \gamma_s p_{rs}(C_1, C_2, ..., C_k) + \notag\\
    &p_C(C_1, C_2, ..., C_{k}).
    \label{eq:twin-vertex-disguised-hamiltonian}
\end{align}

Comparing the expressions for $\freeFermionHamiltonian$ and $\noncontextualHamiltonian$, it is evident that the set of twin-free-fermion-disguised Hamiltonians contains and is not contained by the set of non-contextual Hamiltonians. In this section, we prove that all non-contextual Pauli Hamiltonians are twin-free-fermion-disguised, though not necessarily line graphs. 

\begin{theorem}[Non-Contextual Pauli Hamiltonians Are Twin-Free-Fermion Disguised]\label{thm:NCareFFS}
    Every non-contextual Pauli Hamiltonian $H$ admits a mapping to a twin-free-fermion-disguised Hamiltonian: the frustration graph $\mathcal{G}$ of $H$ always becomes a line graph after projecting out all its twin vertices (i.e., retaining only one vertex from each set of twin vertices). Furthermore, non-contextual Hamiltonians form a strict subset of twin-free-fermion-disguised Hamiltonians with four or more anticommuting operators.
\end{theorem}

\begin{proof}
    The proof follows from direct inspection of the classes of Hamiltonians: we present a graph-based proof as an alternative. \\ 
    \indent For a non-contextual Pauli Hamiltonian $H$ generated with respect to a set of anticommuting Paulis $\mathcal{A}$ and a commuting group $G$, the frustration graphs can be characterized as follows. All connected components of the frustration graphs of these Hamiltonians can be defined by $l$ sets of vertices $S_1, S_2, ..., S_l$ such that all vertices within $S_r$ are disconnected, but all vertices in $S_r$ are connected to all other vertices in all $S_s$. Note that this vertex structure characterizes Paulis such that within each $S_r$ all terms commute, but anticommute with all other terms. A graph with this structure is often called {\it complete $l$-partite} \cite{Ralli_ContextPauli}. \\ 
    \indent Generally, these graphs are not line graphs. However, the vertices in $S_r$ are twins with one another, i.e., share the same set of neighbor vertices. These twin vertices can be projected out of the Hamiltonian and removed from the frustration graph such that each set of vertices $S_r$ contracts down to a single representative vertex. Removing twin vertex symmetries from all sets of vertices $S_r$ results in a complete graph $K_l$, and the complete graph is a line graph. Thus, all non-contextual Pauli Hamiltonians {have a mapping to free-fermions in disguise}. \\ 
    \indent Furthermore, let there exist two anticommuting elements $\gamma_p$ and $\gamma_q$, from sets $S_p$ and $S_q$ respectively. 
    The product of the two elements is, by definition, the product of two anticommuting operators and Paulis which commute with both anticommuting elements. Adding this new vertex to the frustration graph, the modified frustration graph is no longer complete multipartite, and thus the corresponding Hamiltonian is no longer non-contextual. However, the frustration graph is still a line graph augmented with twin vertices, and the Hamiltonian is twin-free-fermion-disguised. We note that we need at least four operators for this to truly be {\it strict} inclusion: for only $3$ anticommuting operators, all free fermion Hamiltonians are also non-contextual. \\
    A {\it Lean} formalization of this theorem can be found in \href{https://github.com/Whitfield-Group-QC/FreeFermionNonContextual_Paper_Lean}{this GitHub repository} \cite{githubLean}.
\end{proof}


\subsection{Hardness of Ground State Solving and Time Evolution of Non-Contextual and Free-Fermion Hamiltonians}\label{sec:hardness}

\indent For a brief review of definitions in complexity theory, see Appendix~\ref{app:qcomp}. Non-contextual Hamiltonians inherit worst-case ground-state energy hardness from the Ising Hamiltonians. Ising Hamiltonians are contained by non-contextual Hamiltonians, so the non-contextual ground state problem is $\mathrm{NP}$-hard \cite{Kirby_ContextClassic}. While we cannot, in general, find the ground state efficiently there may be an efficient way to diagonalize the Hamiltonian or find a unitary which does \cite{patel2023extensionexactlysolvablehamiltoniansusing}. To diagonalize non-contextual Hamiltonians and attempt to perform their Hamiltonian simulation, we utilize a theorem about combinations of solvable Hamiltonians.

\begin{theorem}[Patel, Yen, and Izmaylov 2023, Combinations of Solvable Hamiltonians~\cite{patel2023extensionexactlysolvablehamiltoniansusing}]
    Let $\{H_v\}$ be a family of Hamiltonians diagonalized by $U_v = e^{X_v}$ for anti-Hermitian $X_v$, and let $\{P_v\}$ denote a set of complete orthogonal projectors such that for all $v,v'$, $[H_v, P_{v'}] = 0$ and $[X_v, P_v] = 0$. Then $\sum_v H_v P_v$ is diagonalized by the unitary 
    \begin{equation}
        U = \sum_v U_v P_v = e^{\sum X_v P_v}.
        \label{eq:projector-diagonalization-unitary}
    \end{equation} 
\end{theorem}
We connect this fact to our work by expressing $\{X_v\}$ with operators in terms of specified Majoranas and their products, and $\{P_v\}$ projectors over the mutual eigenspace of Abelian generators. Abelian Pauli symmetries of the Hamiltonian can be tracked efficiently via {\it qubit tapering}, a procedure which uses a Clifford to represent all Abelian Pauli symmetries as Pauli $\{Z_i\}$ with support over a separate register from the non-symmetry terms \cite{bravyi2017taperingqubitssimulatefermionic}. The projectors $\{P_v\}$ under tapering are mapped to projectors over the $2^k$ bitstrings for which the commuting $Z$s act non-trivially on. \\

\indent This allows us to express these Hamiltonians as sums of solvable Hamiltonians, each with an associated projector. This is consistent with the fact that the non-contextual ground state problem is known to be $\mathrm{NP}$-complete, requiring optimizing over exponentially many ground state values for a solution \cite{Kirby_ContextClassic, Ralli_ContextPauli}. The exponential dependence on the dimension of the stabilizer subspace also proves that Trotter-error-free time evolution is exponentially costly in worst case. \\
\indent Finding the ground state of a twin-free-fermion-disguised Hamiltonian, and performing its Hamiltonian simulation without error, are inefficient for the same reasons as for the non-contextual Hamiltonians. 
The worst-case complexity of the ground-state energy problem may be $\mathrm{NP}$, even for $\freeFermionHamiltonian$.\\
\indent For non-contextual and twin-free-fermion-disguised Hamiltonians, evolution to Trotter error $\epsilon$ can be performed via a sequence of simulations of anticommuting and free-fermion Hamiltonians respectively. For non-contextual Hamiltonians, the multipartite frustration graphs can be broken up into fragments of cliques, and the time evolution can be approximated
\begin{equation}
    e^{-\mathrm{i} \noncontextualHamiltonian t} \approx e^{-\mathrm{i} \anticommutingHamiltonian^1 t} e^{-\mathrm{i} \anticommutingHamiltonian^2 t}...
    \label{eq:noncontextual-product-formula}
\end{equation}
Likewise, the twin-free-fermion-disguised Hamiltonians can be broken up into fragments of line graphs, each twin vertex associated with a different line graph. Time evolution can thus be naturally approximated 
\begin{equation}
    e^{-\mathrm{i} \freeFermionHamiltonian t} \approx e^{-\mathrm{i} \lineGraphFreeFermionHamiltonian^1 t} e^{-\mathrm{i} \lineGraphFreeFermionHamiltonian^2 t}...
    \label{eq:free-fermion-product-formula}
\end{equation}

\section{Non-Contextual Fermionic Hamiltonians}
\label{sec:fermionic-encodings}
\indent In this section we derive non-contextual fermionic Hamiltonians from their qubit counterparts, and show that non-contextual Pauli Hamiltonians generate partial fermion-to-qubit encodings.

\begin{theorem}[Non-Contextual Pauli Hamiltonians and Fermionic Encodings]
    Every non-contextual $n$-qubit Pauli Hamiltonian specifies a (family of) fermionic encodings. Furthermore, with respect to Dirac fermion operators, non-contextual Hamiltonians can be written 
     
\begin{equation}
\begin{split}
H = {} & \sum_{{j=1}}^{l} \Big[ a_{{j}} {\big(p_{{2j-1}}{(n_{l+1}, n_{l+2}, ..., n_n)}} \\
& \qquad {{} - \mathrm{i}p_{{2j}}{(n_{l+1}, n_{l+2}, ..., n_n)}\big)} \\
& \quad + a_{{j}}^{\dagger} {\big(p_{{2j-1}}{(n_{l+1}, n_{l+2}, ..., n_n)}} \\
& \qquad {{} + \mathrm{i}p_{{2j}}{(n_{l+1}, n_{l+2}, ..., n_n)}\big)} \Big] \\
& + {P_N p_N}{(n_{l+1}, n_{l+2}, ..., n_n)} \\
& + p_C{(n_{l+1}, n_{l+2}, ..., n_n)}
\end{split}
\label{eq:dependent-fermionic-noncontextual-hamiltonian}
\end{equation}
    with number operators $\{n_i\}$, ${P_N} = {(\Id - 2n_{1}) (\Id - 2n_{2})...(\Id - 2n_{{l}})}$, the parity operator over {the first $l$} modes{, where $l = \lfloor 
    \kLinIndepAnti/2 \rfloor$ for $\kLinIndepAnti$ being the number of independent anticommuting Paulis $\{\gamma_k\}$ of the Hamiltonian, and $p_N = 0$ when the number of anticommuting Paulis $\{\gamma_k\}$ is even}. 
\end{theorem}

\begin{proof}
Let us begin with a non-contextual Hamiltonian 
    \begin{equation}
    \begin{split}
    \noncontextualHamiltonian = {} & \sum_{{r}} \gamma_{{r}} p_{{r}} {(C_{1}, C_{2}, ..., C_{\kLinIndepComm})} \\
    & + p_C{(C_{1}, C_{2}, ..., C_{\kLinIndepComm})}
    \end{split}
    \notag
    \label{eq:noncontextual-hamiltonian-encoding}
\end{equation}
with $\{C_j\}$ being {the $\kLinIndepComm$ independent commuting generators} defined in \eqref{eq:noncontextual-hamiltonian}{, which commute with each other and with every $\gamma_r$}.
    Identifying a fermion-to-qubit transform within a non-contextual Hamiltonian comes from specifying $\kLinIndepAnti$ pairwise anticommuting Paulis $\{\gamma_{{r}}\}$ as Majorana operators, and determining if and how to extend this set of operators to a maximally anticommuting set of Paulis, and thus a fermion-to-qubit encoding. {Throughout, $\kLinIndepAnti$ denotes the number of independent specified Majorana operators $\gamma_1, \dots, \gamma_\kLinIndepAnti$, and $l = \lfloor \kLinIndepAnti/2 \rfloor$ the number of fermionic modes they define.} \\

    \indent There are two scenarios: $Q=\gamma_1\gamma_2...\gamma_{\kLinIndepAnti} =\prod_{{r}} \gamma_{{r}} {\propto} \Id$ and $Q
    {\not\propto} \Id$. \\ 

    \indent { \textbf{ Case 1, $\mathbf{\prod_{{r}} \gamma_{{r}} {\propto} \Id}$:}} We can express the set of ${\kLinIndepAnti+1 = 2l+1}$ {specified} Majorana operators as 
    \begin{equation}
        \{ \gamma_1, \gamma_2, ..., \gamma_{{r}}, \gamma_{1}\gamma_2 ...\gamma_{{\kLinIndepAnti}}\} \notag
        \label{eq:dependent-majorana-set}
    \end{equation}
    by removing the designated parity operator, and assuming $l < n$, the truncated set of ${\kLinIndepAnti = 2l}$ operators can be extended to a set of $2n$ {\it independent} anticommuting operators. Let the operators 
    \begin{equation}
        \mathbf{C} = \{\tilde{\gamma}_{{\kLinIndepAnti+1}}, \tilde{\gamma}_{{\kLinIndepAnti+2}}, ...,  \tilde{\gamma}_{2n}\} \notag
        \label{eq:dependent-majorana-extension}
    \end{equation}
    be the operators which extend the original encoding{, chosen as described under \textit{Choice of extension} below}. A perfect matching of terms in $\mathbf{C}$, 
    \begin{equation}
        M(\mathbf{C}) = \{\tilde{\gamma}_{{\kLinIndepAnti+1}} \tilde{\gamma}_{{\kLinIndepAnti+2}},...,\tilde{\gamma}_{2n-1}\tilde{\gamma}_{2n} \} \notag
        \label{eq:dependent-majorana-matching}
    \end{equation}
    {generates a maximal Abelian group of} symmetries of the specified Majorana operators {\it and} their parity operator, and {for this choice of extension it contains the commuting generators $C_1, \dots, C_k$} in the non-contextual Pauli Hamiltonian. \\
    \indent { \textbf{Case 2, $\mathbf{\prod_{{r}} \gamma_{{r}} {\not\propto} \Id}$:}} If $\prod_{{r}} \gamma_{{r}} {\not\propto} \Id$, it implies the set of specified Majorana operators is a linearly independent set \cite{sarkar2019setscommutinganticommutingpaulis}. {If $\kLinIndepAnti = 2l$ is even, we} extend this set of $\kLinIndepAnti$ Majorana operators to a complete independent anticommuting set of $2n${, chosen as in Case 1,}
    \begin{equation}
        \mathbf{C} = \{\tilde{\gamma}_{{2l+1}}, \tilde{\gamma}_{{2l+2}}, ...,  \tilde{\gamma}_{2n}\} \notag
        \label{eq:independent-majorana-extension}
    \end{equation}
{and no parity operator appears. If $\kLinIndepAnti = 2l+1$ is odd, the product $Q \propto \gamma_1 \cdots \gamma_{\kLinIndepAnti}$ commutes with every $\gamma_k$ (each $\gamma_k$ anticommutes with an even number of the others) and with every $C_j$, and $Q \not\propto \Id$. Then $\gamma_r \propto (\gamma_1 \cdots \gamma_{2l})\, Q$: we let $\gamma_1, \dots, \gamma_{2l}$ define the $l$ modes, treat $Q$ as an additional commuting symmetry alongside $C_1, \dots, C_m$, and extend $\{\gamma_1, \dots, \gamma_{2l}\}$ to $\mathbf{C}$ as above. Case 1 is the special case $Q \propto \Id$.} \\
\indent {\textbf{Choice of extension.} The extension must be chosen with respect to the commuting generators. The operators $\{\gamma_1, \dots, \gamma_{2l}\}$ and the commuting generators $\{C_j\}$ (together with $Q$ when $\kLinIndepAnti$ is odd) have the same commutation relations as the Jordan--Wigner Majorana operators, so there is a Clifford unitary $U$ mapping the latter to the former \cite{gottesmanknill, Aaronson_2004}. Taking $\tilde{\gamma}_r = U \gamma_r U^{\dagger}$ for $k > 2l$, each $C_j$ (and $Q$) is, up to sign, the matched pair $-\mathrm{i}\tilde{\gamma}_{2l+2j-1}\tilde{\gamma}_{2l+2j} = U Z_{l+j} U^{\dagger}$.} \\
\indent For the non-contextual Majorana fermionic Hamiltonian, it is sufficient to replace {each} commuting Pauli $C_{{j}}$ {by the matched pair} ${-\mathrm{i}}\tilde{\gamma}_{{2l+2j-1}} \tilde{\gamma}_{{2l+2j}}$ {to which it is equal up to sign} and write 
\begin{align}
    \noncontextualHamiltonian &= \sum_{{r}} \gamma_{{r}} p_{{r}} {(\tilde{\gamma}_{2l+1}\tilde{\gamma}_{2l+2}, \tilde{\gamma}_{2l+3} \tilde{\gamma}_{2l+4}, ..., \tilde{\gamma}_{2n-1} \tilde{\gamma}_{2n})} \notag\\
        & + p_C{(\tilde{\gamma}_{2l+1}\tilde{\gamma}_{2l+2}, \tilde{\gamma}_{2l+3} \tilde{\gamma}_{2l+4}, .., \tilde{\gamma}_{2n-1} \tilde{\gamma}_{2n})} \notag
    \label{eq:majorana-noncontextual-hamiltonian}
\end{align}
\indent The expression of non-contextual Hamiltonians in terms of Dirac fermionic operators is much less compact, but can be expressed by carefully mapping the Majorana operators indexed by ${2j-1}$ and ${2j}$ to $a_{{j}} + a_{{j}}^{\dagger}$ and $\mathrm{i}({a_{{j}}^{\dagger} - a_{{j}}})$. For the specified Majorana operators ${\{\gamma_k\}}$ we do direct substitution: for the products of the extension Majorana operators, we utilize the identity ${-\mathrm{i}}\tilde{\gamma}_{{2j-1}} \tilde{\gamma}_{{2j}} = {\Id - 2n_{{j}}}$, to re-express the polynomials $ p_{{k}} {(\tilde{\gamma}_{2l+1}\tilde{\gamma}_{2l+2}, \tilde{\gamma}_{2l+3} \tilde{\gamma}_{2l+4}, .., \tilde{\gamma}_{2n-1} \tilde{\gamma}_{2n})}$ as
$ p_{{k}} {(n_{l+1}, n_{l+2}, ..., n_n)}$. Making substitutions, we recover Eq.~\eqref{eq:dependent-fermionic-noncontextual-hamiltonian}. In case 1, ${l = \kLinIndepAnti/2}$, and ${P_N}$ is the parity operator on modes $1$ to ${l}${, which arises from $\gamma_{2l+1} \propto \gamma_1 \cdots \gamma_{2l} = \mathrm{i}^{l} P_N$}{. The parity of modes $l+1$ to $n$ is a polynomial in $n_{l+1}, \dots, n_n$ and is absorbed into $p_N$}. \\
\indent In case 2, {the same expression holds with $l = \lfloor \kLinIndepAnti/2 \rfloor$. If $\kLinIndepAnti$ is even, no parity term arises and $p_N = 0$. If $\kLinIndepAnti$ is odd, $\gamma_{\kLinIndepAnti} \propto (\gamma_1 \cdots \gamma_{2l})\, Q \propto P_N\, Q$, where by the choice of extension $Q$ is up to sign a product of factors $\Id - 2n_j$ with $j > l$; the $\gamma_{\kLinIndepAnti}$ term is therefore of the form $P_N p_N(n_{l+1}, \dots, n_n)$, with $p_N$ containing the factor representing $Q$. When $Q$ is not generated by $C_1, \dots, C_m$, it accounts for one number operator among $n_{l+1}, \dots, n_n$ beyond those needed for the $C_j$.} \\
\indent A {\it Lean} formalization of this theorem can be found in \href{https://github.com/Whitfield-Group-QC/FreeFermionNonContextual_Paper_Lean}{this GitHub repository} \cite{githubLean}.

\end{proof}

\section{Non-Contextuality and Free Fermions in Molecular Hamiltonians}
\label{sec:molecular-hamiltonians}

\indent Contextual decompositions and extensions via fermionic insights have shown promise in solving problems in quantum chemistry, namely contextual subspace variational quantum algorithms, specifically for finding molecular ground states \cite{Kirby_ContextVQE, Ralli_ContextVQE, Weaving_ContextVQE}. The general strategy is to find subsets of the operators in molecular Hamiltonians which are non-contextual, to serve as classical approximations for the ground state. 

\begin{figure}
    \centering
    \includegraphics[width=0.9\linewidth]{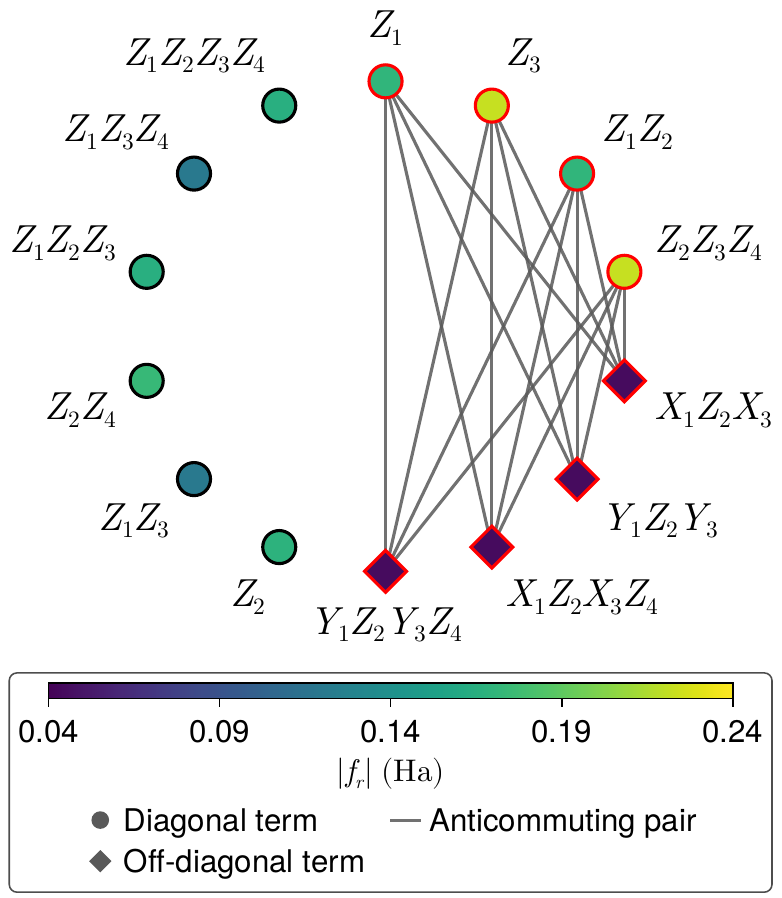}
    \caption{The frustration graph of H$_{2}$ Hamiltonian discretized by STO-3G under the Bravyi--Kitaev encoding with interleaved spin-orbital ordering. Each vertex is a Pauli term $O_r$ of the Hamiltonian (the identity term is omitted), labeled by its corresponding Pauli-operator. Two vertices are joined by an edge when their terms anticommute. The fill color of a vertex labels the magnitude $|f_r|$ of its term coefficient, and its shape labels the term type: circles mark diagonal terms, which contain only $\Id$ and $Z$, and diamonds mark off-diagonal terms, which contain at least an $X$ or $Y$. The eight vertices outlined in red form the complete bipartite graph $K_{4,4}$, whose two partite sets (in two shapes respectively) are each a set of twins, i.e., vertices with identical neighborhoods.}
    \label{fig:H2frus}
\end{figure}

We review an example in which the molecular Hamiltonian is already non-contextual: the H$_2$ Hamiltonian from the minimal STO-3G basis, 
written as a qubit Hamiltonian with the Bravyi--Kitaev encoding \cite{Hempel_2018}. Specifically, we show how this example induces an incomplete fermionic encoding, and is twin-free-fermion-disguised {\it not} via the Bravyi--Kitaev encoding, despite being written with respect to it. \\ 
\indent The H$_2$ Hamiltonian, expressed with the Bravyi--Kitaev encoding, can be written generally as 

\begin{equation*}
\begin{aligned}
H^{BK} =& f_0\Id + f_1 Z_2 + f_2 Z_1 Z_2 Z_3 Z_4 + f_3Z_2Z_4 +  f_4Z_1Z_3 + \\
 &f_5Z_1Z_2Z_3 + f_6 Z_1Z_3Z_4 + f_7 Z_1Z_2 + \\
     & f_8 Z_3 + f_9 Z_1 + f_{10}Z_2Z_3Z_4 + f_{11} X_1 Z_2 X_3 + \\
     &f_{12} Y_1 Z_2 Y_3 + f_{13} X_1 Z_2 X_3 Z_4 + f_{14} Y_1 Z_2 Y_3 Z_4\\
\end{aligned}
\label{eq:h2-bravyi-kitaev-hamiltonian}
\end{equation*}
for real coefficients $f_r$. The term ordering has been defined to make use of contextual and fermionic identities. This Hamiltonian is non-contextual; this can be seen by decomposing it into three sets of Pauli operators 
\begin{align*}
    G &= \{\Id, Z_2, Z_1 Z_2 Z_3 Z_4,Z_2Z_4,Z_1Z_3,Z_1Z_2Z_3, Z_1Z_3Z_4 \}
    \\
    S_1 &= \{Z_1Z_2, Z_3, Z_1, Z_2Z_3Z_4  \}
    \\
    S_2 &= \{X_1 Z_2 X_3 , Y_1 Z_2 Y_3, X_1 Z_2 X_3 Z_4, Y_1 Z_2 Y_3 Z_4\}
\end{align*}
We now demonstrate two facts: this set of operators is induced by a partial encoding, and the frustration graph is that of a twin-free-fermion-disguised Hamiltonian. \\ 
\indent From $S_1$ and $S_2$ we choose representatives $\gamma_1 = Z_1$ and $\gamma_2 = X_1 Z_2 X_3 Z_4$ respectively (thus, $k'=2$, and the extension below contains $2n-k'=6$ Majoranas). This partial encoding is $\{Z_1, X_1 Z_2 X_3 Z_4 \}$. There are many extensions of this encoding; one is 
\begin{align*}
    \{\tilde{\gamma}_r \} = &\{{Y_1Z_2Z_4, -X_1Z_2Z_3, Y_1Z_2Z_3X_4,} \notag\\
    & {Y_1Z_2Z_3Y_4, Y_1Y_2Y_3Z_4, -Y_1X_2Y_3Z_4}\}
\end{align*}
Let the pairing of Majorana operators in the extension be 
\begin{align*}
    M(\{\tilde{\gamma}_r\}) = \{{-\mathrm{i}}\tilde{\gamma}_1 \tilde{\gamma}_2, {-\mathrm{i}}\tilde{\gamma}_3 \tilde{\gamma}_4, {-\mathrm{i}}\tilde{\gamma}_5 \tilde{\gamma}_6 \} = \{Z_1Z_3Z_4, Z_4, Z_2 \}
\end{align*}
 $M(\{\tilde{\gamma}_r\})$ and products of terms in $M(\{\tilde{\gamma}_r\})$ generate the set $G$. We can rewrite the elements of $S_1$ and $S_2$ in terms of the Majorana operators and products of operators in $M(\{\tilde{\gamma}_r\})$  (up to signs), 
\begin{align*}
    S_1 &= \{\gamma_1 \tilde{\gamma}_5\tilde{\gamma}_6, \gamma_1 \tilde{\gamma}_1\tilde{\gamma}_2\tilde{\gamma}_3\tilde{\gamma}_4, \gamma_1, \gamma_1 \tilde{\gamma}_1\tilde{\gamma}_2\tilde{\gamma}_5\tilde{\gamma}_6  \}
    \\
    S_2 &= \{\gamma_2 \tilde{\gamma}_3 \tilde{\gamma}_4, \gamma_2\tilde{\gamma}_1\tilde{\gamma}_2, \gamma_2, \gamma_2 \tilde{\gamma}_1\tilde{\gamma}_2\tilde{\gamma}_3\tilde{\gamma}_4 \}
\end{align*}
We have thus expressed all operators in the Hamiltonian with respect to a partial fermionic encoding, and the induced twin vertex symmetries of the partial encoding. \\
\indent This Hamiltonian is also twin-free-fermion-disguised. The frustration graph can be found in FIG.~\ref{fig:H2frus}, and the twin-vertex symmetries can be seen both graphically and in the structure of sets of vertices induced by groupings $S_1$ and $S_2$.


\begin{figure}[h]
    \centering
    \includegraphics[width=0.9\linewidth]{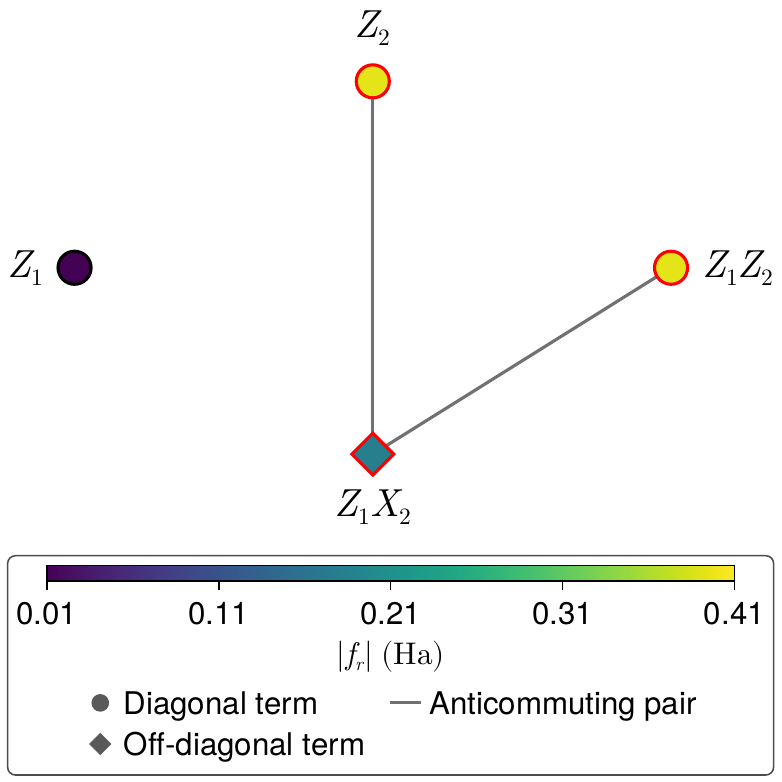}
    \caption{The frustration graph of H$_{2}$ Hamiltonian discretized by STO-3G under the Bravyi--Kitaev encoding in the particle-number sector where the numbers of alpha-spin electrons and beta-spin electrons are both one (the identity term is omitted). This graph is a line graph with two connected components, where the non-isolated vertices (outlined in red) form an induced line graph of a path graph on four vertices.}
    \label{fig:H2frus_sec11}
\end{figure}

To study non-contextuality and fermionic solvability in molecular Hamiltonians, it appears we have three main control factors: molecular geometry, basis set, and fermionic encoding. However, the choice of encoding is not important when examining spin solvability embedded in an electronic Hamiltonian. This is highlighted through a corollary related to our H$_2$ example. \\
\begin{corollary}[Contextuality and Encoding in Molecular Hamiltonians]
    For any molecular Hamiltonian, discretized in any basis set, whether it (or any fragment of it) is non-contextual or twin-free-fermion-disguised does not depend on which Jordan--Wigner-like fermion-to-qubit encoding is used. 
\end{corollary}
The H$_2$ Hamiltonian discretized with STO-3G is non-contextual and twin-free-fermion-disguised in every such encoding. We can show this equivalence between non-contextuality and free-fermion solvability for H$_2$ by applying qubit tapering based on $\mathcal{Z}_2$ symmetry to its Hamiltonian to reach the fixed particle-number sector. Under the minimal STO-3G discretization, forcing the parity of orbitals in both spin sectors to be odd is equivalent to setting the particle number to one in each spin sector. As a result, the frustration graph of this subspace Hamiltonian (with one alpha-spin electron and one beta-spin electron) is trivially a line graph, as shown in FIG.~\ref{fig:H2frus_sec11}.

As discussed in Sec.~\ref{sec:hardness}, although the time evolution of both non-contextual and free-fermion Hamiltonians 
can be efficiently simulated with bounded error, solving for the ground states remains hard in the worst-case scenarios. However, for a free-fermion-solvable Hamiltonian whose frustration graph is one single connected component, the ground-state problem is also efficiently solvable provided there are no non-identity cycle symmetries. Hence, in the next two subsections, we investigate the practicality of finding the largest free-fermion-solvable connected components, using heuristic approaches based on either non-contextuality or frustration line graphs. Specifically, we are interested in both the size and scale of these connected components.

Thus, the objective is to find connected components with the highest possible coefficient $\ell_1$ norms and compare them with other fragment types in Hamiltonian simulation problems. \\ 

\subsection{Non-Contextual Fragments}



In general, molecular Hamiltonians are contextual and not free-fermion solvable. Therefore, 
we investigate the practicality of heuristically searching for the maximal free-fermion-solvable Hamiltonian fragments. 
For a given Hamiltonian $H=\sum_{r\in V}\coeff_rO_r$ where $\coeff_r$ is the real coefficient for the Pauli-operator term $O_r$, we define its {\it fragment} as
\begin{equation}
    H_F=\sum_{r\in F}\coeff_rO_r
\end{equation}
where $F\subseteq V$ is a subset of all the Pauli-operator terms $V$ of the (encoded) Hamiltonian $H$ determined by a search algorithm such that $H_F$ satisfies a well-defined property, e.g., non-contextuality. We then quantify the free-fermion solvability of realistic molecular systems beyond H$_2$ by two metrics: the count ratio $|F|/|V|$ and the coefficient $\ell_1$-norm ratio $\lone{F}/\lone{V}$ (i.e., the fragment-coefficient norm over the Hamiltonian-coefficient norm). In particular, we deliberately exclude identity terms from $V$ so they contribute to neither metric.


\begin{table*}[htbp]
\centering
\renewcommand{\arraystretch}{1.1}
\setlength{\tabcolsep}{4pt}
\begin{tabular}{l r r r r r}
\toprule
System & $n$ & $|V|$ & $\pooldiag$ & $\poolquad$ & Connected $\poolquad$ \\
\midrule
$\mathrm{H_3^{+}}$ &  6 &   51 &  21 (0.412, 0.791) &  8 (0.157, 0.247) &  4 (0.078, 0.092) \\
$\mathrm{H_4}$     &  8 &  104 &  36 (0.346, 0.570) & 12 (0.115, 0.092) &  4 (0.038, 0.027) \\
$\mathrm{LiH}$     & 12 &  630 &  78 (0.124, 0.794) & 36 (0.057, 0.375) & 16 (0.025, 0.150) \\
$\mathrm{H_2O}$    & 14 & 1085 & 105 (0.097, 0.773) & 42 (0.039, 0.612) & 16 (0.015, 0.249) \\
\bottomrule
\end{tabular}
\caption{
Structural information of the tested qubit Hamiltonian instances from the data set of contextual molecular systems (spin-singlet, discretized by STO-3G) under the Jordan--Wigner encoding, provided by the Symmer library. For the headers, $n$ is the number of qubits required for the encoding (i.e., the number of spin orbitals for each system), $|V|$ is the number of vertices in each frustration graph. $\pooldiag$, $\poolquad$, and Connected $\poolquad$ represent the diagonal block, the quadratic sector, and a maximum-size connected component of the frustration
graph induced on $\poolquad$, respectively.  Each entry starting from the fourth column reads: ``member count, (count ratio, coefficient $\ell_1$-norm ratio)'', with the identity term excluded from every $\ell_1$ norm. The number of vertices $|V|$ is equal to the number of (non-zero-coefficient) terms in the corresponding Hamiltonian, excluding the identity term. }
\label{tab:sys-db-1body}
\end{table*}

Our demonstrative test samples for finding the maximal free-fermion solvable fragment consists of four contextual molecular Hamiltonians provided by the open-source software library, Symmer~\cite{githubSymmertestshamiltonian_dataMain}: spin-singlet H$_3^+$, H$_4$, LiH, and H$_2$O in the STO-3G basis, under the Jordan--Wigner encoding, ranging from $n\!=\!6$ to $14$ qubits and $|V|=51$ to $1085$ terms. The structural information for these Hamiltonians is listed in Table~\ref{tab:sys-db-1body}, which also includes two types of fragments that are trivially free-fermion solvable. 

Before considering free-fermion or non-contextual decompositions, it is worth singling out two fragments we can always find in a molecular Hamiltonian. The first type of fragment is the diagonal block $\pooldiag$. It collects the terms whose underlying Majorana operators are adjacent Majorana pairs. Such a term represents a product of Majorana pairs $\gamma_{2p-1}\gamma_{2p}\propto 2n_p-\Id$, hence diagonal in the occupation-number basis, and under the Jordan--Wigner encoding it is a operator of $Z$'s. All of these terms commute with one another, so $\pooldiag$ is a commuting Hamiltonian in its own right: its frustration graph has no edges, it is trivially non-contextual, and its time evolution is a product of commuting exponentials. It also carries most of the coefficient weight of a molecular Hamiltonian. The diagonal block contains only 10--41\% of the terms but 57--79\% of the coefficient $\ell_1$-norm ratio, reflecting the dominance of Hartree--Fock molecular orbital energies ($h_{pp}$) and diagonal Coulomb repulsions ($V_{pqqp}$) in the molecular Hamiltonian.

The second type of fragment is the quadratic sector $\poolquad$. Its terms are products of two Majorana operators with coefficients $h_{pq}$, which are free-fermion solvable regardless of the underlying encoding. This fragment has some overlap with the diagonal block, through the diagonal quadratic terms. The frustration graph of this fragment is the line graph of a graph whose vertices are the $2n$ Majorana operators, since two quadratic Majorana operators anticommute when they share one Majorana. However, this line graph need not be connected. 
Furthermore, for a free-fermion-solvable fragment to be meaningful in a realistic molecular Hamiltonian, it should extend beyond quadratic terms to include (two-body) quartic terms. Otherwise, the mean-field methods become useless for approximating electron-electron interactions in molecular systems. We could always, for example, diagonalize the free-fermion sector so the only free-fermion terms were already encapsulated in the diagonal block. 

\begin{table*}[htbp]
\centering
\renewcommand{\arraystretch}{1.1}
\setlength{\tabcolsep}{4pt}
\begin{tabular}{l r *{4}{w{r}{85pt}}}
\toprule
System & $|V|$ & NC (by count) & CNC (by count) & NC (by coefficient) & CNC (by coefficient) \\
\midrule
$\mathrm{H_3^{+}}$ &   51 &  27 (0.529, 0.818) & 16 (0.314, 0.421) &  25 (0.490, 0.831) & 16 (0.314, 0.495) \\
$\mathrm{H_4}$     &  104 &  50 (0.481, 0.452) & 28 (0.269, 0.231) &  40 (0.385, 0.630) & 24 (0.231, 0.328) \\
$\mathrm{LiH}$     &  630 & 100 (0.159, 0.808) & 44 (0.070, 0.268) & 100 (0.159, 0.808) & 44 (0.070, 0.268) \\
$\mathrm{H_2O}$    & 1085 & 131 (0.121, 0.796) & 52 (0.048, 0.104) & 131 (0.121, 0.796) & 52 (0.048, 0.104) \\
\bottomrule
\end{tabular}
\caption{The non-contextual groupings of the four
molecular systems detailed in Table~\ref{tab:sys-db-1body}, obtained via Symmer's depth-first-search heuristic grouping algorithm (i.e., \texttt{NoncontextualOp} without early stop by a time constraint). There are two search strategies provided for this algorithm:
\texttt{DFS\_largest} and \texttt{DFS\_magnitude}, which specify whether the grouping prioritizes the number of terms or their coefficient $\ell_1$ norms. Both strategies were tested and their results are listed in the table, labeled by ``(by count)'' and ``(by coefficient)'', respectively. Furthermore, for a given strategy, a pair of related groupings are recorded (in the same entry format as columns 4--6 in Table~\ref{tab:sys-db-1body}): NC represents the default output grouping from the algorithm, the maximal non-contextual terms; CNC represents the (locally) maximal connected component of NC.}
\label{tab:symmer-noncontextual}
\end{table*}

The first direct search objective we tested is finding the maximal non-contextual fragments. The search algorithm we applied is the one behind the function \texttt{NoncontextualOp} from Symmer (version 0.0.12), a depth-first search over subsets of terms with two primary strategies: \texttt{DFS\_largest} (by count), which prioritizes the number of terms, and \texttt{DFS\_magnitude} (by coefficient), which prioritizes their $\ell_1$ norm. The results are shown in Table~\ref{tab:symmer-noncontextual}. By {Definition}~\ref{id:nc}, a non-contextual (NC) fragment splits into its commuting part $G$, the set of terms that commute with every other term in NC, and the remainder, whose frustration graph is both complete multipartite and connected, i.e., the connected non-contextual fragment (CNC). Consequently, $G$ is equivalently the set of all isolated vertices in NC, and $|G|=|\mathrm{NC}|-|\mathrm{CNC}|$. The table explicitly shows the sizes and the coefficient $\ell_1$-norm ratios of both NC and CNC, suggesting that a non-contextual fragment typically consists mostly of $G$. For H$_2$O, $G$ accounts for 79 of the 131 terms and 69\% of the total norm, compared with 10\% for the CNC; for LiH, it accounts for 56 terms and 54\% of the total norm, compared with 27\%; for the two smaller molecules, the split is roughly even with the CNC slightly heavier, except for H$_3^+$ under the by-coefficient strategy, where the CNC carries noticeably more weight (0.495 versus 0.336). Comparing with Table~\ref{tab:sys-db-1body}, the coefficient $\ell_1$-norm ratio of NC is comparable to that of the diagonal block for the same system (0.808 versus 0.794 for LiH, 0.796 versus 0.773 for H$_2$O), which suggests that the non-contextuality-based search algorithm from Symmer does not provide a significant advantage over the trivially twin-free-fermion-disguised fragment provided by $\pooldiag$. The two strategies return the same fragment for LiH and H$_2$O and differ modestly for H$_3^+$ and H$_4$, so the picture is not an artifact of the search objective.

By Theorem~\ref{thm:NCareFFS} in Sec.~\ref{sec:free-fermion-solvability}, each CNC is twin-free-fermion-disguised: the terms within each equivalence class partitioning the CNC are twin vertices of its frustration graph, so projecting them out contracts every class to its representative $\gamma_r$, and what results is a complete graph $K_l$, the line graph of a star graph $K_{1,l}$. Hence, NC are free-fermion disguised, typically with many twin-vertex symmetries.
For time evolution, the CNC sets the cost, and one may ask whether switching from the complete-multipartite requirement to the line-graph heuristics can help find larger connected free-fermion-solvable fragments.

\subsection{Connected Line-Graph Fragments}\label{sec:clg-test}

\begin{table*}[ht]
\centering
\renewcommand{\arraystretch}{1.1}
\setlength{\tabcolsep}{4pt}
\begin{tabular}{l r *{4}{w{r}{85pt}}}
\toprule
 & & \multicolumn{2}{c}{Ranked by count ($\rankflag=\False$)}
   & \multicolumn{2}{c}{Ranked by $\ell_1$ norm ($\rankflag=\True$)} \\
\cmidrule(lr){3-4} \cmidrule(lr){5-6}
System & $|V|$ & $\ordk$ & $\ordc$ & $\ordk$ & $\ordc$ \\
\midrule
$\mathrm{H_3^{+}}$ &   51 & 16 (0.314, 0.223) & 16 (0.314, 0.269) & 16 (0.314, 0.223) & 16 (0.314, 0.269) \\
$\mathrm{H_4}$     &  104 & 22 (0.212, 0.194) & 22 (0.212, 0.194) & 22 (0.212, 0.194) & 22 (0.212, 0.194) \\
$\mathrm{LiH}$     &  630 & 96 (0.152, 0.320) & 50 (0.079, 0.163) & 96 (0.152, 0.320) & 50 (0.079, 0.163) \\
$\mathrm{H_2O}$    & 1085 & 91 (0.084, 0.523) & 37 (0.034, 0.254) & 91 (0.084, 0.523) & 37 (0.034, 0.254) \\
\bottomrule
\end{tabular}
\caption{The restricted free-fermion solvable groupings of the four
molecular systems detailed in Table~\ref{tab:sys-db-1body}, obtained via various configurations of the encoding-invariant, line-graph-based search specified by Algorithm~\ref{alg:search}. Specifically, the configurations are under the support-descending visiting order $\ordk$
and the coefficient-descending order $\ordc$, ranked by term count ($\rankflag=\False$) or by
coefficient $\ell_1$ norm ($\rankflag=\True$). The resulting grouping for each configuration follows the same format as columns 4--6 in Table~\ref{tab:sys-db-1body}. For the details of this algorithm, refer to Appendix~\ref{app:line-graph-alg}.}
\label{tab:line-graph-search}
\end{table*}

The second search objective we tested is the maximal (frustration) connected line-graph fragments. {By the result} of Chapman and Flammia~\cite{Chapman_Fermions}, such a fragment is quadratic in the Majorana operators of some fermion-to-qubit encoding, and it is diagonalized and can be time-evolved exactly by a free-fermion circuit conjugated by the Clifford that changes the encoding, with no Trotter errors {(within each cycle-symmetry sector, if any)}. Although the frustration line-graph criterion has been briefly benchmarked against contextuality-based criteria~\cite{patel2024exactly}, to the best of our knowledge, no existing literature has presented a detailed description of a robust search algorithm based on the line-graph criterion, as of the writing of this manuscript. Therefore, we developed a deterministic and (locally) optimal heuristic search algorithm to conduct our tests. In particular, the search runs multiple times by starting from a fragment as a subset of the quadratic sector, as shown in Table~\ref{tab:sys-db-1body}, and grows its size to the maximum by adding other terms in the Hamiltonian. Then, the best fragment is selected by a ranking rule based on lexicographic ordering, prioritizing either the count ($\rankflag=\False$) or the coefficient $\ell_1$ norm ($\rankflag=\True$), mirroring Symmer's strategy configurations. More specifically, in each run branch, the search picks a connected component of the quadratic sector as the initial fragment seed, which is a connected line-graph component by construction, and then offers the remaining terms one at a time as candidates. For each candidate term, it is admitted if the fragment's frustration graph remains a connected line graph; it is deferred if it is not yet connected to the fragment; otherwise, it is permanently rejected, since line-graph membership is inherited by induced subgraphs. The deferred terms are later re-offered until no single term can be added, so the fragment is maximal in that sense. Each admission check is a line-graph recognition, linear in the size of the graph \cite{ROUSSOPOULOS1973108}, and the number of checks is polynomial in $|V|$. The output is the connected fragment $\frag^\star$. Terms commuting with every term of $\frag^\star$ could be appended afterward as isolated vertices without breaking solvability, exactly as $G$ pads a non-contextual fragment, so the like-for-like comparison is between $\frag^\star$ and CNC.

\begin{figure*}[htbp]
    \centering
    \includegraphics[width=0.9\linewidth]{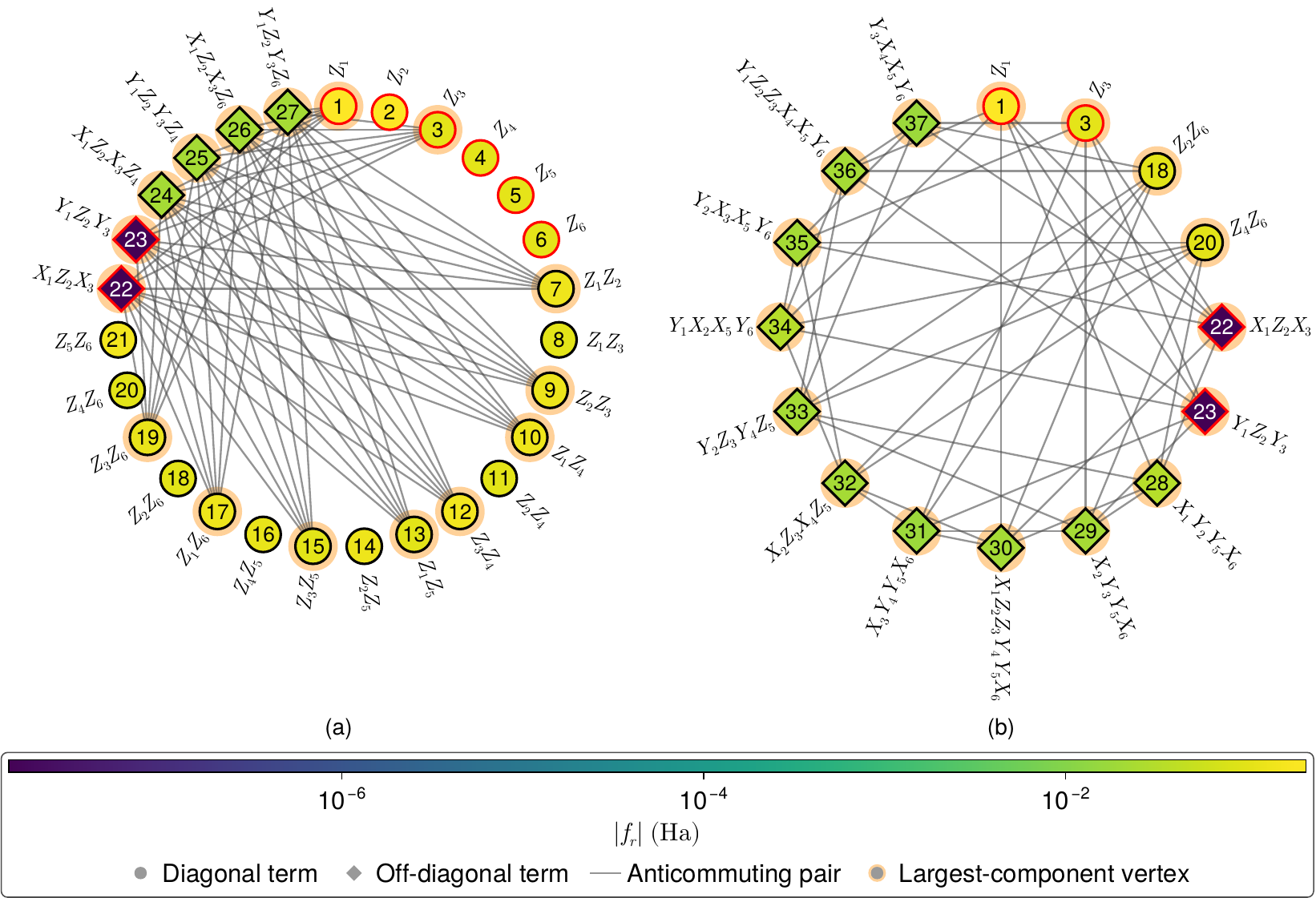}
    \caption{The comparison between two fragments selected from the Hamiltonian of H$_3^+$ in the
STO-3G basis under Jordan--Wigner transformation (six qubits; 51 terms excluding the identity term), each drawn as the frustration graph it induces. Circles are diagonal terms (Pauli operators containing $I$ and $Z$ only),
diamonds are off-diagonal terms (containing $X$ or $Y$), a red outline marks the quadratic
sector, and the fill color
encodes the coefficient magnitude $|f_r|$ in Hartree on the logarithmic scale of the
shared color bar. Orange halos mark the vertices of the largest connected component of
each graph. The vertices are numbered consistently across the panels, a term carrying the
same number wherever it appears. (a)~The non-contextual fragment produced by Symmer's
\texttt{DFS\_largest} search, numbered 1--27 clockwise from the top: all 21 diagonal
terms of the Hamiltonian together with the six off-diagonal terms. These 27 terms decompose into a connected component of 16 terms, isolated from 11 diagonal terms. (b)~The line-graph fragment 
produced by our encoding-invariant search in the support-descending visiting order configuration. This fragment contains 16 connected terms, so the halo covers all the terms. Moreover, its frustration graph is the $4\times4$ rook's graph, the line graph of the complete bipartite graph $K_{4,4}$ drawn in FIG.~\ref{fig:h3p1_clg_root} with the same number labels.}
\label{fig:H3p1_nc_vs_clg}
\end{figure*}

This search algorithm is also encoding-agnostic by using Majorana supports, index sets that label each encoded Pauli operator by its original Majorana operator indices. As a result, the seeds and the visiting order in which the candidate terms are offered are functions of the Majorana supports, which every Jordan--Wigner-like $n$-to-$n$ encoding (e.g., Bravyi--Kitaev encoding) assigns identically to the terms of the same fermionic Hamiltonian. Under the support-descending visiting order $\ordk$ (i.e., a purely lexicographic ordering of the Majorana supports), which consults no coefficient, the fragment is therefore independent of the encoding choice by construction. The coefficient-descending order $\ordc$, which offers the heaviest terms first (and only uses the support order as a tie breaker), is invariant up to the numerical noise in the term coefficients. For more details of this algorithm, we refer the reader to Appendix~\ref{app:line-graph-alg}.

Table~\ref{tab:line-graph-search} lists the maximal connected line-graph fragments found by our search algorithm. Three observations stand out. First, ranking the branches by count or by coefficient $\ell_1$ norm makes no difference: the largest fragment is also the heaviest for all four systems, and in every case the winning branch was seeded by a component of the quadratic sector rather than by the seed grown greedily from the rest of the Hamiltonian. Second, these maximal fragments reach beyond the quadratic sectors. For LiH and H$_2$O, they contain 96 and 91 terms against 16 for the largest connected quadratic component, and more than the entire quadratic sector (36 and 42 terms), so at least 60 and 49 of their terms correspond to the quartic Majorana operators. These fragments are free-fermion models in an induced encoding in which the quartic terms become quadratic, as discussed in Sec.~\ref{sec:free-fermion-solvability} and demonstrated in the H$_2$ example at the beginning of this section. Third, the support-based ordering outperforms the coefficient-based ordering. On LiH and H$_2$O, the former finds fragments twice as large (96 versus 50 terms and 91 versus 37) and twice as heavy (32\% versus 16\% and 52\% versus 25\% of the norm), and only on H$_3^+$ does the coefficient-based ordering gain a slightly higher weight at equal size. Hence, we conclude that accepting coefficient-heavy terms first is generally suboptimal. The heaviest terms are typically from the diagonal block; a diagonal term anticommutes with every term that shares one of its constituent Majorana operators. Admitting such terms early can exclude many of the quartic terms that the support-based ordering admits, via the claw and the other forbidden induced subgraphs for line graphs. Overall, compared to the search results also based on line-graph criteria in work~\cite{patel2024exactly} (labeled as ``FF''), our search algorithm (under $\ordk$) managed to find larger free-fermion-solvable fragments for contextual molecular systems such as LiH and H$_2$O. We note that our results were obtained under stricter constraint of only admitting a single connected line-graph component (i.e., free of pairwise commuting terms) for each tested system. 

Additionally, compared with the CNC found by Symmer's search algorithm in Table~\ref{tab:symmer-noncontextual}, our search algorithm is mostly on par with, if not better, in finding the maximal free-fermion-solvable fragments, depending on the system size (spin-orbital number). For H$_3^+$, both fragments have 16 terms, but the CNC is heavier (42\% or 50\% versus 22\% or 27\% under different search 
strategies); for H$_4$, the CNC is larger (28 and 24 terms versus 22) and heavier (23\% and 33\% versus 19\%). However, for the two larger systems, the best line-graph fragment wins the best CNC fragment on both measures: 96 against 44 terms and 32\% against 27\% of the coefficient $\ell_1$-norm ratio for LiH, and 91 against 52 terms and 52\% against 10\% of the norm for H$_2$O, a fivefold gain in coefficient weight. This significant performance discrepancy reflects distinct levels of anticommutative constraints that the two heuristics impose during the search for free-fermion-solvable fragments. A complete multipartite CNC requires every pair of terms from different commutative equivalence classes to anticommute, which is a global constraint whose capacity is set by the $2n+1$ anticommuting representatives.  In contrast, the line-graph approach asks only that each term anticommute with terms sharing one of its corresponding Majorana operators, a local constraint that matches the sparse, hopping-like anticommutation pattern of Majorana operators in an electronic Hamiltonian. As the Hamiltonian grows, the line-graph fragment continues to grow empirically, whereas the CNC does not.

For ground-state approximation in the spirit of contextual subspace methods, the diagonal-sector coefficient weight is essential, and non-contextual fragments remain the natural tool. However, the two search algorithms technically target different subclasses of the twin-free-fermion-disguised Hamiltonians of Eq.~\eqref{eq:twin-vertex-disguised-hamiltonian}: a CNC is a line graph only after its redundant twin vertices are projected out, while our search demands a line graph without such projection, which is the sense in which the groupings of Table~\ref{tab:line-graph-search} are restricted; these two searches do not find the same local optimal fragment in general. Nevertheless, for simulating the anticommuting part of a molecular Hamiltonian, the results in Tables~\ref{tab:symmer-noncontextual} and~\ref{tab:line-graph-search} suggest that the line-graph-based approach may be a viable alternative approach for searching for maximal free-fermion-solvable fragments in practice, especially for the heavier solvable piece. Additionally, we note that these line-graph results come from a polynomial-time search algorithm. \\
\indent As a final note regarding the connected line graph components found with our algorithms, there {\it are} non-trivial cycle symmetries present, but only a small number of them. $H_3^+, H_4$, and $H_2O$ all have a cycle symmetry group generated by two operators, LiH with 3 generators. For practical application of these fragments for time evolution or ground state solving, the presence of $O(1)$ cycle symmetry generators for an $n$ qubit Hamiltonian doesn't restrict solvability. 

\begin{figure}[htbp]
    \centering
    \includegraphics[width=0.9\linewidth]{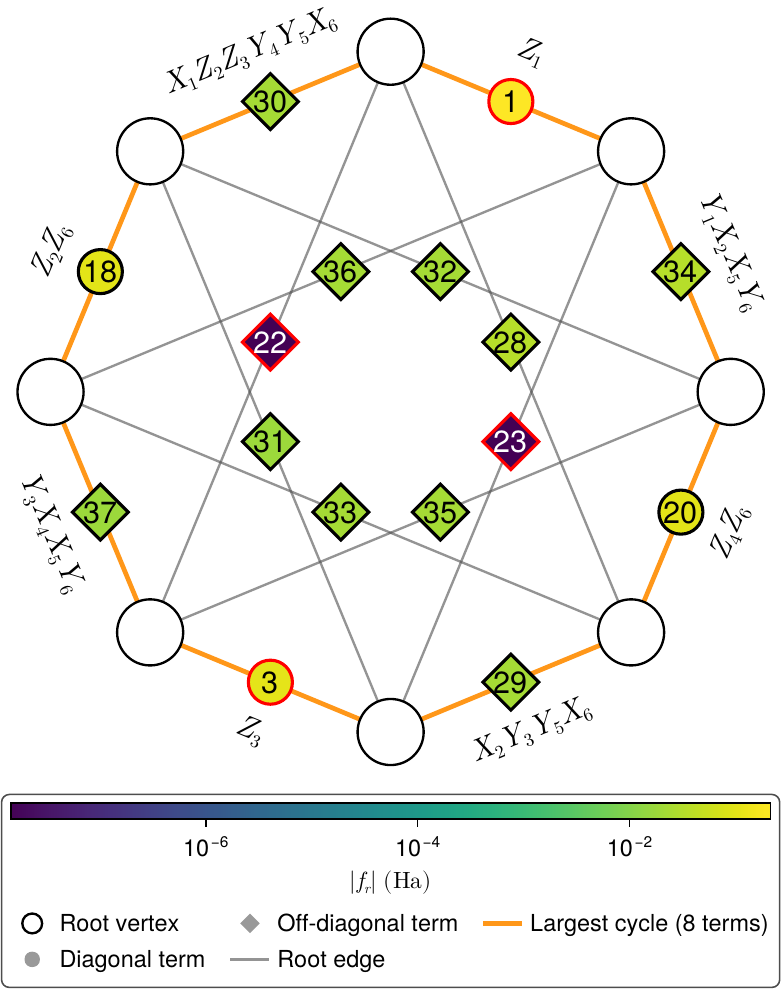}
    \caption{The root graph $\mathcal{R}$ of the fragment in FIG.~\ref{fig:H3p1_nc_vs_clg}(b): every term is an
edge of $\mathcal{R}$, and two terms anticommute exactly when their edges share a
vertex. $\mathcal{R}$ is the complete bipartite graph $K_{4,4}$; its eight vertices
(white circles) lie on a circle, and its 16 edges are the terms, each carrying at its
midpoint the marker it has in FIG.~\ref{fig:H3p1_nc_vs_clg}(b): shape, fill color, number,
and the red outline of the quadratic sector on $Z_1$, $Z_3$, $X_1Z_2X_3$ and
$Y_1Z_2Y_3$. The largest term cycle is laid around the circle in orange.}
\label{fig:h3p1_clg_root}
\end{figure}

\section{Conclusion}
\label{sec:conclusion}

\indent In this paper, we define non-contextual fermionic Hamiltonians from their Pauli relatives, and prove that all non-contextual Pauli Hamiltonians are twin-free-fermion-disguised. We then investigate maximal non-contextual and free-fermion fragments for decomposing molecular Hamiltonians. We find that when considering the connected components of non-contextual and line-graph fragments, the connected line-graph components are comparable in size while offering better complexity-theoretic guarantees of ground-state solvability and Hamiltonian simulation. This extends and clarifies the work of Patel and Izmaylov \cite{patel2024exactly}, and Patel, Yen, and Izmaylov \cite{patel2023extensionexactlysolvablehamiltoniansusing}, adding deeper theoretical analysis while being transparent about graph algorithms and symmetries in line graphs. \\
\indent An open question our work poses is in free-parafermion solvability \cite{Mann_2025} and its relationship to qudit non-contextuality. It is known that contextuality and related quantities like Wigner negativity are easier to study in $d>2$ dimensions \cite{Howard_ContextMagic, Raussendorf_2017, PhysRevA.85.062121, PhysRevA.70.062101}; is there a relationship between free parafermions and qudit contextuality \cite{Mann_2025}? \\
\indent The tested use cases for contextual subspace methods, specifically for molecular Hamiltonians, have been restricted to variational quantum algorithms. Contextual subspace methods can also be used in Hamiltonian simulation, as a strategy for grouping Pauli terms that goes beyond mutually commuting or anticommuting strategies. In particular, we show that these grouping strategies can be viewed at the Majorana level and incorporated into grouping strategies that better leverage induced fermionic structure in spin Hamiltonians. Our work suggests that outside of variational quantum algorithms and eigensolvers, non-contextual Hamiltonians may contain limited utility in quantum information theoretic task \cite{patel2023extensionexactlysolvablehamiltoniansusing, Kirby_ContextVQE, Ralli_ContextVQE, Weaving_ContextVQE}. We thus suggest that more time be spent considering the applicability of free fermion solvability as a technique for finding practical advantage in quantum information.   \\ 
 

\begin{acknowledgments}
\indent AMP thanks Basie Seibert, Sam Alterman, and Sam Elman for insightful discussion. AMP and JDW thank Brent Harrison and Jason Necaise for insightful discussion. AMP and JDW are supported by the Department of Energy under award DE‐SC0019374. WW and JDW are supported by the Army Research Office under award W911NF2410043. KAR is supported by the National Science Foundation Graduate Research Fellowship Program under Grant No. 2236868.
JDW holds concurrent appointments at Dartmouth College and as an Amazon Visiting Academic. This paper describes work performed at Dartmouth College and is not associated with Amazon.
\end{acknowledgments}

\section*{Data Availability Statement}

A copy of this manuscript, its associated {\it Lean} proofs, and data for plots can be found in \href{https://github.com/Whitfield-Group-QC/FreeFermionNonContextual_Paper_Lean}{this GitHub repository}.

\section*{AI Usage disclosure}

AMP used Anthropic Claude Opus 5.0 and Claude Opus 5.5 for comments on notation and consistency throughout the editing of this document. AMP also used Claude Opus 5.5 to generate the Lean code that formalizes Theorems 2 and 4, and reworded the natural-language proof of Theorem 4 for better auto-formalization. WW designed the initial version of the line-graph search algorithm proposed in the paper, then used Claude Fable 5 to help incorporate Majorana supports to achieve the encoding-invariant feature. WW also used Claude Fable 5 to help implement programming scripts for debugging the algorithm and to benchmark it against Symmer's non-contextual algorithms. In the context of typesetting and copy editing, WW used Claude Fable 5.1 to help format the pseudocode listings and to help generate the initial version of the plotting scripts (which were later human-reviewed and revised) for visualizing the frustration graphs (FIG. 3--8) presented in the paper. 

\bibliography{Contextuality}

\appendix 
\section{Quantum Complexity} \label{app:qcomp}

\indent 
Here we briefly review some facts and definitions from the field of complexity theory, and quantum complexity specifically. We are both interested in problems and their instances: both problem statements, and specific inputs to those problems statements. A problem belongs to a complexity class if there exists an algorithm that solves every instance within the class's resource bounds; membership is thus a worst-case upper bound. \\ 
\indent The { most well known complexity classes} are the classical ones of $\mathrm{P}$ and $\mathrm{NP}$. Complexity classes can be formally defined with respect to various computational models, i.e. Turing machines. $\mathrm{P}$ is the class of binary output decision problems which can be solved with a deterministic Turing machine in a polynomial amount of compute time. 

The parameter chosen for this polynomial matters; some example are: the molecular specification as a string e.g. H2O or CH4 for $poly(N_{atoms})$, the number of qubits used $poly(n)$, the number of terms in a decomposition $poly(\#  terms)$, or the number of basis functions used $poly(m)$. 
For the purposes of this paper we assume the problem input, whether we are considering ground-state energy approximation or Hamiltonian simulation, has polynomial size with respect to the number of qubits.




$\mathrm{NP}$ can be defined similarly: it is the class of decision problems which can be solved on a {\it nondeterministic} Turing machine in polynomial compute time. 
Stated in terms of verification, $\mathrm{NP}$ is the class of problems which can be {\it verified} efficiently on a deterministic Turing machine. Important concepts in complexity theory include {\it hardness} and {\it completeness}.  A problem is class-hard if it is known that any problem in the class can be reduced to it, where reduction is an algorithm which transforms one problem to another. A problem is class-complete if it is both hard and contained in that class.



\indent The quantum analogs to the above are the classes $\mathrm{BQP}$ and $\mathrm{QMA}$. The statements of the classes are generalizations of the definitions above, except with the added knowledge of working with a probabilistic computer. $\mathrm{BQP}$ is the class of decision problems solvable on a quantum computer in polynomial time, with an error probability at most $1/3$ for all instances. This error can be made arbitrarily small with repeated application of the measurement procedure and the use of the Chernoff bound \cite{kolmogorov1933foundations}. \\
\indent To make connection with the ground Hamiltonians we have defined and will define throughout the rest of this paper, we rephrase the ground state problem for a Hamiltonian as a decision problem. However, it is a decision problem with a promise: not simply a yes/no choice, but a yes/no choice dependent on some promise. 
\begin{identity}[Hamiltonian Ground-State Decision Problem]
    Given a Hamiltonian $H$ with $\| H\|\leq 1$ in the operator norm, given constants $\alpha > \beta$ with
    { \it promise gap}  $\alpha - \beta \geq 1/q(n)$ for some polynomial $q$
    in the input size $n$, and given promise that ground-state energy
    $E_0(H)$ satisfies either
    \[
        E_0(H) \leq \beta
        \qquad\text{or}\qquad
        E_0(H) \geq \alpha,
    \]
    decide whether $E_0(H) \leq \beta$ (YES) or
    $E_0(H) \geq \alpha$ (NO).
\end{identity}
The stabilizer, anticommuting, and free-fermion Hamiltonian ground state problems belong to the class $\mathrm{P}$: there exist algorithms which run in polynomial size of the input which can solve for the ground states for these Hamiltonians. The commuting Hamiltonian belongs to the class $\mathrm{NP}$. The commuting Hamiltonians contain the family of classical Ising Hamiltonians, which are known to be $\mathrm{NP}$-complete \cite{FBarahona_1982, Lucas_2014, Whitfield_2012}. \\
\indent The interacting electronic Hamiltonian ground state problem is in the class $\mathrm{QMA}$ \cite{Whitfield_QMA, ogorman}. \\
\indent We are interested not only in ground state solving, but performing time evolution of the Hamiltonian. 

{\begin{identity}[Hamiltonian Simulation Decision Problem]
    Given an $n$-qubit Hamiltonian $H=\sum_{p=1}^{m}H_p$ with
    $m,\;t,\;\max_p\|H_p\|=\mathrm{poly}(n)$, where each $e^{-iH_p s}$
    admits an efficient quantum circuit, a state $|\psi\rangle$, and
    a designated outcome $0$ on the first qubit, let
    \[
        p=\big\|(|0\rangle\langle 0|\otimes \Id)\,e^{-iHt}|\psi\rangle\big\|^{2}_2.
    \]
    Promised that either $p\geq\tfrac{2}{3}$ (YES) or $p\leq\tfrac{1}{3}$ (NO),
    decide which inequality is true.
\end{identity}
 Hamiltonian simulation is often performed via Trotterization and product formulas. The first-order Trotter approximation can be written
\begin{align*}
    e^{-iHt}\;&\approx\; U_r:=\Big(\prod_{p=1}^{m}e^{-iH_p t/r}\Big)^{r},
    \\[4pt]
    \epsilon \leq \big\|e^{-iHt}-U_r\big\|
    \;&\leq\; \frac{t^{2}}{2r}\sum_{p<q}\big\|[H_p,H_q]\big\|
    \\[4pt]
    \;&\leq\; \frac{(m\Lambda t)^{2}}{r},
    \qquad \Lambda=\max_p\|H_p\|.
\end{align*}
Performing $r=O\!\big((m\Lambda t)^{2}/\epsilon\big)=\mathrm{poly}(n,1/\epsilon)$
Trotter steps gives error at most $\epsilon$, where epsilon is a function of the number of terms, the operator norm of terms, $t$, and $r$ \cite{Lloyd1996UniversalQS}. Applying $U_r$ to $|\psi\rangle$, measuring, and
taking a majority vote over $O(1)$ repetitions decides the problem in polynomial time. 
}

Hamiltonian simulation is $\mathrm{BQP}$-complete \cite{Low_2019}.  For our work, we are interested not in whether some family can simulate others, but how hard it is on a quantum computer to perform that simulation. Take, as an example, an Ising Hamiltonian. Solving for its ground state is in $\mathrm{NP}$. 
Its time evolution $e^{-\mathrm{i}Ht}$ can though, be implemented on a quantum computer with no error $\epsilon$ in a polynomial number of gates even on simple nearest neighbor geometries. In this paper we will make a distinction between cases in which we can perform error free evolution, versus when we cannot. \\ 
\indent Before proceeding, we make a final note: throughout this paper, we will not consider Hamiltonians composed of an exponential number of operators. The commuting Hamiltonians, for example, could have a maximum $2^n$ number of pairwise commuting elements over $n$ qubits. This restriction confines Ising-like and commuting Hamiltonians to only a polynomial number of terms, and guarantees that, for example, we are not considering how to do the ``efficient'' exponentiation of an exponentially scaling number of Paulis.  \\

\clearpage
\onecolumngrid 
\section{Encoding-Invariant Frustration Line-Graph Search} \label{app:line-graph-alg}

This appendix includes the pseudocode for our encoding-invariant line-graph search algorithm (the main algorithm presented in Algorithm~\ref{alg:search}). This algorithm is benchmarked against Symmer's non-contextual fragment search algorithm in Sec.~\ref{sec:clg-test}. The main algorithm executes several sub-algorithms, listed separately. First, Algorithm~\ref{alg:supports} constructs Majorana supports, labels of Majorana operators that are invariant under different ($n$-to-$n$) fermionic-encoding representations. Second, Algorithm~\ref{alg:expand} performs a single-pass extension of a connected line-graph component, given a sequence of Pauli-operator terms as candidates. Third, Algorithm~\ref{alg:order} initializes the visiting order of a given set (or subset) of Hamiltonian terms based on different heuristic configurations. We use the term ``sequence'' to refer to an ordered set whose elements can be indexed, and $\CLG$ denotes the class of connected line graphs.

\begin{algorithm}[htbp]
\caption{Majorana supports of the terms in an $n$-mode encoded Hamiltonian $H$. Any Majorana operator $\gamma_q$ (represented by a Pauli operator) anticommutes with $\prod_{q'\in\supp}\gamma_{q'}$ when $q\in\supp$ and $|\supp|$ is even, or $q\notin\supp$ and $|\supp|$ is odd. All Majorana operators that anticommute with a Pauli-operator term $O_r$ therefore form an (index-set) support or a support complement of $O_r$, depending on the parity of those Majorana operators.}
\label{alg:supports}
\begin{algorithmic}[1]
\Require Terms $\{O_r\}$ ($r\in V$) in $H$; Majorana operators $\{\gamma_q\}$ ($1\leq q \leq 2n$) of the applied fermionic encoding as Pauli operators.
\Ensure Majorana supports $\{\supp_r\}$ with $\supp_r\subseteq\{1,\dots,2n\}$ such that
        $O_r\propto\prod_{q\in\supp_r}\gamma_q$ (i.e., equality up to a phase factor) for every $r\in V$.
\Procedure{Supports}{$\{O_r\}_{r\in V}$, $\{\gamma_q\}_{q=1}^{2n}$}
  \ForAll{$r\in V$}
    \State $\supp_r\gets\{\,q : \ac{O_r}{\gamma_q}=0\,\}$
    \If{$|\supp_r|$ is odd} $\supp_r\gets\{1,\dots,2n\}\setminus\supp_r$
    \EndIf
  \EndFor
  \State \Return $\{\supp_r\}_{r\in V}$
\EndProcedure
\end{algorithmic}
\end{algorithm}

\begin{algorithm}[htbp]
\caption{A greedy extension of a connected line-graph fragment $F$ by appending candidates from a sequence of terms $B$ in a single-pass visit. The admitted candidates must not break the line-graph property of extended $F$. Specifically, a candidate term with no frustration edge into (the non-empty) $F$ is deferred to a pool $D$ so that it stays eligible for potential future passes; a candidate whose admission would break the line-graph property is rejected permanently, because such a property is hereditary under induced subgraphs.
}
\label{alg:expand}
\begin{algorithmic}[1]
\Require a fragment $\frag$ of terms (as a sequence) such that
        $\mathcal{G}[\frag]\in\CLG$; a
        sequence $\cand$ of candidate terms with $\frag\cap\cand=\varnothing$;
        the terms $\{O_r\}$ associated with $F$ and $B$ are assumed accessible without being explicit inputs.
\Ensure  $\frag$ extended in acceptance order, and the deferred candidates $\defer$ in visiting order.
\Procedure{Expand}{$\frag$, $\cand$}
  \State $\defer\gets()$ \Comment{initialize $D$ as an empty sequence}
  \ForAll{$r\in\cand$, in order}
    \If{$\frag\neq()$ and $\comm{O_r}{O_s}=0\;\;\forall s\in\frag$}
      \State append $r$ to $\defer$
           \Comment{deferred}
    \ElsIf{$\mathcal{G}[\frag\cup\{r\}]\in\CLG$}
      \State append $r$ to $\frag$
           \Comment{accepted}
    \Else
      \State discard $r$
           \Comment{rejected permanently}
    \EndIf
  \EndFor
  \State \Return $(\frag,\,\defer)$
\EndProcedure
\end{algorithmic}
\end{algorithm}

\begin{algorithm}[htbp]
\caption{Visiting order (sequence) of an input set of Pauli-operator terms $X$. The base order $\ordk$ ($\coefflag=\False$) is a lexicographical descending order of integer-key representations $\skey$ of Majorana supports for the terms, which consults no term coefficients. When $\coefflag=\True$, the visiting order changes to order $\ordc$: the magnitudes of term coefficients become the primary sorting keys and the base order is only used to break their exact ties. The integer key for the support of term $O_r$ is generated via mapping  $\skey(r)=\bigl(|\supp_r|,\;\;\sum_{p:\,2p-1\in\supp_r}2^{\,p-1}+\sum_{p:\,2p\in\supp_r}2^{\,n+p-1}\bigr)$.}
\label{alg:order}
\begin{algorithmic}[1]
\Require a set of terms $X\subseteq V$; order flag $\coefflag\in\{\False,\True\}$; objects used without being explicit inputs: Majorana supports $\{\supp_r\}$, term coefficients $\{\coeff_r\}$.
\Ensure $X'$: the reordered visiting sequence for $X$.
\Procedure{Order}{$X$, $\coefflag$}
  \State $X' \gets$ list $X$ in descending order of $\skey$ 
  \Comment{for each $r \in X$ associated with $O_r$, use $\skey(r)$ as the sorting key}
  \If{$\coefflag=\True$} re-sort $X'$ by descending $|\coeff_r|$
  \EndIf
  \State \Return $X'$
\EndProcedure
\end{algorithmic}
\end{algorithm}

\begin{algorithm*}[htbp]
\caption{An encoding-invariant search for a maximally connected line-graph fragment $\frag^\star$, given an encoded Hamiltonian $H$. Pools, seeds, and visiting orders all depend only on the Majorana supports; the term coefficients affect $\frag^\star$ only through Algorithm~\ref{alg:order} when $\coefflag=\True$ and through the final ranking. Each seed $\seed_w$ induces a connected line-graph fragment, either through the line-graph heredity of the quadratic sector $\poolquad$ or through Algorithm~\ref{alg:expand}, hence initiating a greedy search branch.
Each branch runs two passes of Algorithm~\ref{alg:expand} and keeps re-offering previously deferred terms until the fragment reaches a fixed point, where no single term of $H$ can be added to it. The ranking flag $\rankflag$ re-sorts the converged branch fragments before picking an optimal fragment as the returned value. Additionally, $\ell_1(\cdot)$ returns the input's coefficient $\ell_1$ norm; $\keyseq(\cdot)$ returns the input's support keys $\skey(r)$ in lexicographically ascending order, where $\skey$ is defined in Algorithm~\ref{alg:order}}
\label{alg:search}
\begin{algorithmic}[1]
\Require $H=\sum_{r\in V}\coeff_rO_r$ on $n$ qubits, carrying no identity term and no repeated terms; the Majorana operators $\{\gamma_1,\dots,\gamma_{2n}\}$ of a one-to-one encoding; the flags $\coefflag$, $\rankflag \in\{\False,\True\}$ (coefficient order, ranking by $\ell_1$ norm, fixed-point sweep)
\Ensure the winning fragment $\frag^\star\subseteq V$, its branch label $w^\star$, the states of its four stages, and its support fingerprint
\Procedure{Search}{$H$, $\{\gamma_q\}_{q=1}^{2n}$, $\coefflag$, $\rankflag$}
    \State $\{\supp_r\}_{r\in V}\gets$ \Fn{Supports}{$\{O_r\}_{r\in V}$, $\{\gamma_q\}_{q=1}^{2n}$}
        \Comment{Algorithm~\ref{alg:supports}: construct Majorana supports}
    \State $\pooldiag\gets\{\,r\in V : 2p-1\in\supp_r\Leftrightarrow 2p\in\supp_r,\;\forall p\,\}$
        \Comment{build diagonal block $\pooldiag$ from index-paired supports}
    \State $\poolquad\gets\{\,r\in V : |\supp_r|=2\,\}$
        \Comment{build quadratic sector $\poolquad$}
    \State $\poolrest\gets V\setminus(\pooldiag\cup\poolquad)$
        \Comment{register the residual terms $\poolrest$}
    \vspace{1em}
    \State sort $\pooldiag$, $\poolquad$, and $\poolrest$ in $\skey$-ascending order
        \Comment{$\skey$ defined in Algorithm~\ref{alg:order} is used to generate sorting keys}
    \State $(\seed_1,\dots,\seed_K)\gets$ obtain $K$ connected components of $\mathcal{G}[\poolquad]$, in the order of their appearances in $\poolquad$
    \State $(\seed_{K+1},\defer_{\subrest})\gets$ \Fn{Expand}{$()$, \Fn{Order}{$\poolrest$, $\coefflag$}}
        \Comment{Algorithms~\ref{alg:expand} and~\ref{alg:order}: form a seed pool from $\poolrest$}
    
    \vspace{1em}
    \For{$w=1,\dots,K+1$}
        \Comment{one branch per seed $\seed_w$}
        \If{$w\le K$} 
            \Comment{form branch pools}
            \State $\wpool_1\gets\poolrest$; $\quad\wpool_2\gets\poolquad\setminus\seed_w$
        \Else
            \State $\wpool_1\gets\poolquad$; $\quad\wpool_2\gets\defer_{\subrest}$
        \EndIf

        \State $(\frag^{(w)}_1,\defer^{(w)}_1)\gets$ \Fn{Expand}{$\seed_w$, \Fn{Order}{$(\wpool_1\cup\pooldiag)\setminus\seed_w$, $\coefflag$}}
             \Comment{pass 1}
        \State $(\frag^{(w)}_2,\defer^{(w)}_2)\gets$ \Fn{Expand}{$\frag^{(w)}_1$, \Fn{Order}{$\wpool_2\setminus\frag^{(w)}_1$, $\coefflag$}}
             \Comment{pass 2}
        \State $\frag^{(w)}\gets\frag^{(w)}_2$

        \vspace{1em}
            \State $\cand\gets$ \Fn{Order}{$(\defer^{(w)}_1\cup\defer^{(w)}_2)\setminus\frag^{(w)}$, $\coefflag$}
                \Comment{form the re-offering pool}
            \While{$\cand\neq()$}
                \Comment{repeatedly re-offer terms until the $\frag^{(w)}$ grows to a fixed point}
                \State $(\frag',\cand)\gets$ \Fn{Expand}{$\frag^{(w)}$, $\cand$}
                \State \textbf{if} $|\frag'|=|\frag^{(w)}|$ \textbf{then break}
                    \Comment{self-consistent fixed point is achieved}
                \State $\frag^{(w)}\gets\frag'$
            \EndWhile

    \EndFor
    \vspace{1em}

    \For{$w=1,\dots,K+1$}
        \If{$\rankflag=\True$}
            \Comment{rank (primarily) by term-coefficient $\ell_1$ norm}
            \State $\rankkey(w)\gets\bigl(\lone{\frag^{(w)}},\ |\frag^{(w)}|,\ \keyseq(\frag^{(w)})\bigr)$
        \Else
            \Comment{rank (primarily) by term count}
            \State $\rankkey(w)\gets\bigl(|\frag^{(w)}|,\ \lone{\frag^{(w)}},\ \keyseq(\frag^{(w)})\bigr)$
        \EndIf
    \EndFor

    \State $w^\star\gets$ find the smallest $w$ maximizing $\rankkey(w)$ lexicographically
    \State \Return $\frag^{(w^\star)}$
        \Comment{as the locally optimal fragment $\frag^\star$}
    
\EndProcedure
\end{algorithmic}
\end{algorithm*}

\clearpage

\end{document}